\documentclass[10pt]{article}
\usepackage{bm,latexsym,amsmath,amsthm,amssymb}
\usepackage{graphicx,indentfirst}
\usepackage[numbers,sort&compress]{natbib}
\date{}
\begin{document}
\title{\bf Well-posedness and peakons for a higher-order $\mu$-Camassa-Holm equation}
\author{Feng Wang \thanks{wangfeng@xidian.edu.cn}\\
\small School of Mathematics and Statistics, Xidian University,
    Xi'an 710071, PR China\\[3pt]
Fengquan Li \thanks{fqli@dlut.edu.cn}\\
\small School of Mathematical Sciences, Dalian University of Technology,
Dalian 116024, PR China\\[3pt]
Zhijun Qiao\thanks{zhijun.qiao@utrgv.edu}\\
\small School of Mathematical \& Statistical Sciences,\\
\small University of Texas-Rio Grande Valley, Texas 78539, USA}
\date{}
\maketitle \baselineskip 3pt
\begin{center}
\begin{minipage}{130mm}
{{\bf Abstract.} In this paper, we study the Cauchy problem of a higher-order $\mu$-Camassa-Holm equation.
By employing the Green's function of $(\mu-\partial_{x}^{2})^{-2}$, we obtain the explicit formula of the inverse function $(\mu-\partial_{x}^{2})^{-2}w$ and local well-posedness for the equation in Sobolev spaces $H^{s}(\mathbb{S})$, $s>\frac{7}{2}$. Then we prove the existence of global strong solutions and weak solutions. Moreover, we show that
the data-to-solution map is H\"{o}lder continuous in $H^{s}(\mathbb{S})$, $s\geq 4$, equipped with the $H^{r}(\mathbb{S})$-topology for $0\leq r<s$. Finally, the equation is shown to admit single peakon solutions which have continuous second derivatives and jump discontinuities in the third derivatives.

\vskip 0.2cm{\bf Keywords:} Higher-order $\mu$-Camassa-Holm equation; Global existence; Weak solutions; H\"{o}lder continuous; Peakon solutions.}

\vskip 0.2cm{\bf AMS subject classifications (2000):} 35G25, 35L05, 35B30.
\end{minipage}
\end{center}

\baselineskip=15pt

\section{Introduction}
\label{intro}

In this paper, we discuss the Cauchy problem of the following equation
$$
\begin{array}{l}
m_{t}+2mu_{x}+um_{x}=0,\quad m:=(\mu-\partial_{x}^{2})^{2}u=(\mu+\partial_{x}^{4})u,
\end{array}
\eqno(1.1)
$$
where $u(t,x)$ is a time-dependent but spatially periodic function on the unit-circle $\mathbb{S}=\mathbb{R}/\mathbb{Z}$ and $\mu(u)=\int_{\mathbb{S}}udx$ denotes its mean.
(1.1) was proposed by Escher et al. \cite{eskol14, eswun12} as an Euler equation on diffeomorphism group $\textmd{Diff}^{\infty}(\mathbb{S})$ with respect to certain metric, in which $m$ is defined to be $m=(\mu+(-\partial_{x}^{2})^{r/2})u$, and it is a $\mu$-version of the following higher-order Camassa-Holm equation with $k=2$:
$$
\begin{array}{l}
m_{t}+2mu_{x}+um_{x}=0,\quad m=(1-\partial_{x}^{2})^{k}u,
\end{array}
\eqno(1.2)
$$
which is derived by McLachlan and Zhang \cite{mczh09} as the Euler-Poincar\'{e} differential equation on the Bott-Virasoro group with respect to the $H^{k}$ metric.
In \cite{mczh09}, they proved the global existence of strong solutions and weak solutions. The asymptotic behavior of (1.2) was studied in \cite{mczh11}. Non-uniform dependence on initial data for (1.2) with $k=2$ was showed in \cite{fliu13}. Well-posedness of (1.2) with $k=2$ in Besov spaces was investigated in \cite{tangl15}.
A generalized form of (1.2) was presented in \cite{mzz11}.
When $m=\sum_{j=0}^{k}(-1)^{j}\partial_{x}^{2j}u$, (1.2) becomes into another higher-order Camassa-Holm equation, which was first proposed by Constantin and Kolev \cite{ck03}, and it admits global weak solutions \cite{chk09}.

When $m=(\mu-\partial_{x}^{2})u$, (1.1) is reduced to the following $\mu$-Camassa-Holm equation lying mid-way between the periodic Hunter-Saxton and Camassa-Holm equations (see equations (1.4) and (1.5) later)
$$
\begin{array}{l}
m_{t}+2mu_{x}+m_{x}u=0, \quad m=(\mu-\partial_{x}^{2})u,
\end{array}
\eqno(1.3)
$$
which describes the propagation of weakly nonlinear orientation waves in a massive nematic liquid crystal with external magnetic filed and self-interaction \cite{klm08}. Moreover, (1.3) is also an Euler equation on $\textmd{Diff}^{\infty}(\mathbb{S})$ and it gives rise to the geodesic flow on $\textmd{Diff}^{\infty}(\mathbb{S})$ with the right-invariant metric
at the identity defined by the inner product \cite{klm08}
$$
\begin{array}{l}
\langle f,g\rangle_{\mu}=\mu(f)\mu(g)+\int_{\mathbb{S}}f^{\prime}(x)g^{\prime}(x)dx~.
\end{array}
$$
In \cite{klm08,lmt10}, the authors showed that (1.3) is bi-Hamiltonian and admits both cusped and smooth travelling wave solutions which are natural candidates for solitons. The orbit stability of periodic peakons was studied in \cite{clel13}.

Equation (1.3) is a $\mu$-version of the classical Camassa-Holm equation
$$
\begin{array}{l}
m_{t}+2mu_{x}+m_{x}u=0, \quad m=(1-\partial_{x}^{2})u,
\end{array}
\eqno(1.4)
$$
which was introduced to model the unidirectional propagation of shallow water waves over a flat bottom \cite{ch93}.
(1.4) is another expressional form of the geodesic flow both on the diffeomorphism group of the circle \cite{ck02} and on the Bott-Virasoro group \cite{k07}. It has a bi-Hamiltonian structure \cite{ff81} and is completely integrable \cite{ch93, c98}. Moreover, (1.4) has been extended to an entire integrable hierarchy including both negative and positive flows and shown to admit algebro-geometric solutions on a symplectic submanifold \cite{Qiao2003}. When $m=-\partial_{x}^{2}u$,
(1.4) becomes into the Hunter-Saxton equation,
$$
\begin{array}{l}
m_{t}+2mu_{x}+m_{x}u=0, \quad m=-\partial_{x}^{2}u,
\end{array}
\eqno(1.5)
$$
which was derived in \cite{hs91} as a model for propagation of orientation waves in a massive nematic liquid crystal director field. It also has a bi-Hamiltonian structure \cite{bss01,hs91,or96} and is completely integrable as a special member in the negative flows of Harry-Dym hierarchy \cite{Qiao1998}.
Alternatively, (1.5) can be regarded as the high-frequency or short-wave limit
$(x, t)\mapsto(\epsilon x,\epsilon t)$, for $\epsilon\rightarrow 0$, of (1.4) \cite{dp98}. In recent years, the Cauchy problem of (1.4) and (1.5), in particular its well-posedness, blow-up behavior and global existence, have been well-studied both on the real line and on the circle, e.g., \cite{brc05,brc07,c97,ca00,c05,ce98,d01,himkm10, holli10, hz94,lo00,xz00,y04}.

The present paper is devoted to
studying the existence of global strong solutions of (1.1) in $H^{s}(\mathbb{S})$, $s>\frac{7}{2}$ and global weak solutions in $H^{2}(\mathbb{S})$,
 establishing the H\"{o}lder continuity of the data-to-solution map in $H^{s}(\mathbb{S})$, $s\geq 4$, and proving the existence of single peakon solutions which have continuous second derivatives and jump discontinuities in the third derivatives.

We first give the Green's function of the operator
$(\mu-\partial_{x}^{2})^{-2}$, which plays a key role in writing the explicit formula of the inverse function $(\mu-\partial_{x}^{2})^{-2}w$ and constructing the peaked solutions,
and local well-posedness of (1.1).
Then we show the global existence of strong solutions to (1.1).
Next, for any $T_{0}>0$ and $s\geq 4$, we prove that the data-to-solution map is H\"{o}lder continuous in $H^{s}(\mathbb{S})$, equipped with the $H^{r}(\mathbb{S})$-topology for $0\leq r<s$. Motivated by the recent work \cite{cokar15}, we establish the existence of global weak solution in $H^{2}(\mathbb{S})$ without using an Ole\u{\i}nik-type estimate
(see \cite{chkar06,xz00}),
which is not easy to be verified in numerical experiment. Inspired by the forms of periodic peakons for the $\mu$-Camassa-Holm equation and its modified forms
\cite{lmt10,qfliu14,qfl14}, we show that (1.1) admits the peaked periodic-one traveling-wave solutions. From the results obtained in this paper, one can see that the higher-order $\mu$-Camassa-Holm equation has significant difference from the $\mu$-Camassa-Holm equation, e.g., it does not admit finite time blowup solutions.

The rest of the paper is organized as follows. In Section 2, the Green's function of the operator
$(\mu-\partial_{x}^{2})^{-2}$, the explicit formula of the inverse function $(\mu-\partial_{x}^{2})^{-2}w$ and local well-posedness for (1.1)
with initial data in $H^{s}(\mathbb{S}), s>\frac{7}{2}$, are provided.
In Section 3, we show the global existence of strong solutions. The H\"{o}lder continuity of the data-to-solution map is established in Section 4. In Section 5, we prove the global existence of weak solutions. The existence of single peakon solutions is revealed in Section 6.

\section{Preliminaries}

In this section, we will give the Green's function of the operator
$\Lambda_{\mu}^{-4}:=(\mu-\partial_{x}^{2})^{-2}=(\mu+\partial_{x}^{4})^{-1}$
and establish the local well-posedness for (1.1).

\subsection{Green's Function}

To write the inverse $v=\Lambda_{\mu}^{-4}w$ explicitly and construct the peaked solutions, we need to investigate the Green's function of the operator $\Lambda_{\mu}^{-4}$. We denote the Fourier transform of $f$ by $\widehat{f}$.

For a periodic function $g$ on the circle $\mathbb{S}=\mathbb{R}/\mathbb{Z}$, we have
$$
\widehat{\mu(g)}(k)
=\int_{\mathbb{S}}\mu(g)(x)e^{-2\pi ikx}dx
=\mu(g)\int_{\mathbb{S}}e^{-2\pi ikx}dx
=\mu(g)\delta_{0}(k),
$$
where
$$
\begin{array}{l}
\delta_{0}(k)=\left\{\begin{array}{l}
1, \quad k=0, \\[3pt]
0, \quad k\neq 0.
\end{array}
\right.
\end{array}
$$
Since $\mu(g)=\widehat{g}(0)$, we have $\widehat{\mu(g)}(k)=\delta_{0}(k)\widehat{g}(k)$.
Thus,
$$
\widehat{\Lambda_{\mu}^{4}g}(k)
=\widehat{(\mu+\partial_{x}^{4})g}(k)
=[\delta_{0}(k)+(-2\pi ik)^{4}]\widehat{g}(k).
$$

If $g$ is the Green's function of the operator $\Lambda_{\mu}^{-4}$, that is, $g$ satisfies $\Lambda_{\mu}^{4}g=\delta(x)$,
then $[\delta_{0}(k)+(-2\pi ik)^{4}]\widehat{g}(k)=1$.
Thus,
$$
g(x)
=\sum_{k\in \mathbb{Z}}\widehat{g}(k)e^{2\pi ikx}
=\sum_{k\in \mathbb{Z}}\frac{1}{\delta_{0}(k)+(2\pi k)^{4}}e^{2\pi ikx}
=1+2\sum_{k=1}^{\infty}\frac{\cos 2\pi kx}{(2\pi k)^{4}}.
$$

From Dirichlet's criterion, we know that the series
$$
\sum_{k=1}^{\infty}\frac{\cos 2\pi kx}{(2\pi k)^{4}}, \quad
-\sum_{k=1}^{\infty}\frac{\sin 2\pi kx}{(2\pi k)^{3}}, \quad
-\sum_{k=1}^{\infty}\frac{\cos 2\pi kx}{(2\pi k)^{2}}, \quad
\sum_{k=1}^{\infty}\frac{\sin 2\pi kx}{2\pi k}
$$
converge for any $x\in [0,1)\simeq\mathbb{S}$, and uniformly converge in any closed interval $[\alpha, \beta]\subset (0, 1)$. We may assume that
$$
F(x)=\sum_{k=1}^{\infty}\frac{\cos 2\pi kx}{(2\pi k)^{4}}, \quad x\in [0,1)\simeq\mathbb{S},
$$
then $F(x)$ is three-times continuously differentiable on $[\alpha, \beta]\subset (0, 1)$ and
$$
F^{\prime}(x)=-\sum_{k=1}^{\infty}\frac{\sin 2\pi kx}{(2\pi k)^{3}}, \quad
F^{\prime\prime}(x)=-\sum_{k=1}^{\infty}\frac{\cos 2\pi kx}{(2\pi k)^{2}}, \quad
F^{\prime\prime\prime}(x)=\sum_{k=1}^{\infty}\frac{\sin 2\pi kx}{2\pi k}.
$$

On the other hand, we know
$$
\begin{array}{l}
\sum_{k=1}^{\infty}\frac{\sin 2\pi kx}{2\pi k}
=\left\{\begin{array}{l}
\frac{1}{4}-\frac{x}{2}, \quad 0<x<1, \\[3pt]
0, \quad x=0,
\end{array}
\right.
\end{array}
$$
then there exists a constant $a$ such that
$$
F^{\prime\prime}(x)=-\frac{1}{4}x^{2}+\frac{1}{4}x+a, \quad x\in (0, 1).
$$
Using the fact $\sum_{k=1}^{\infty}\frac{(-1)^{k-1}}{k^{2}}=\frac{\pi^{2}}{12}$ (see Page 194 in \cite{graf08}), we get
$a=F^{\prime\prime}(\frac{1}{2})-\frac{1}{16}
=\frac{1}{4\pi^{2}}\sum_{k=1}^{\infty}\frac{(-1)^{k-1}}{k^{2}}-\frac{1}{16}
=-\frac{1}{24}$.
Thus, some constant $b$ exists such that
$$
F^{\prime}(x)=-\frac{1}{12}x^{3}+\frac{1}{8}x^{2}-\frac{1}{24}x+b, \quad x\in (0, 1),
$$
we can easily deduce that $b=F^{\prime}(\frac{1}{2})=0$. So there exists a constant $c$ such that $$
F(x)=-\frac{1}{48}x^{4}+\frac{1}{24}x^{3}-\frac{1}{48}x^{2}+c, \quad x\in (0, 1).
$$
Using the fact $\sum_{k=1}^{\infty}\frac{(-1)^{k-1}}{k^{4}}=\frac{7\pi^{4}}{720}$, we get
$c=F(\frac{1}{2})+\frac{1}{768}
=-\frac{1}{16\pi^{4}}\sum_{k=1}^{\infty}\frac{(-1)^{k-1}}{k^{4}}+\frac{1}{768}
=\frac{1}{1440}$.

It is easy to check that $F(x)$ is two-times continuously differentiable on $[0,1)\simeq \mathbb{S}$, and so is $g(x)=1+2F(x)$.
Thus, the Green's function $g(x)$ is given by
$$
g(x)=1+2F(x)
=-\frac{1}{24}x^{2}(x-1)^{2}+\frac{721}{720}, \quad x\in [0,1)\simeq \mathbb{S},
$$
and is extended periodically to the real line. In other words,
$$
g(x-x^{\prime})
=-\frac{1}{24}[(x-x^{\prime})^{4}-2|x-x^{\prime}|^{3}+(x-x^{\prime})^{2}]+\frac{721}{720},
\quad x, x^{\prime}\in [0,1)\simeq \mathbb{S}.
$$
Note that $\mu(g)=1$. The graph of $g$ can be seen in Figure 1.
\begin{figure}
\begin{center}
\includegraphics[width=2.50in,height=1.6in]{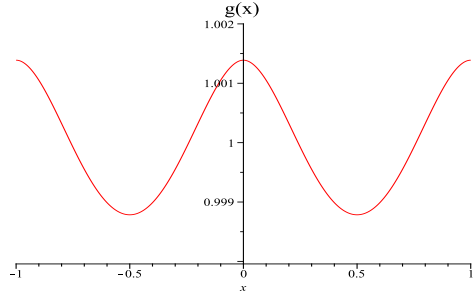}
\caption{The periodic Green's function $g(x)$ corresponding to the operator $\Lambda_{\mu}^{-4}$.}
\end{center}
\end{figure}

The inverse $v=\Lambda_{\mu}^{-4}w$ is given explicitly by
$$
\begin{array}{rl}
v(x)
&=(g*w)(x)=\int_{0}^{1}g(x-x^{\prime})w(x^{\prime})dx^{\prime}\\[3pt]
&=\left(-\frac{x^{4}}{24}+\frac{x^{3}}{12}-\frac{x^{2}}{24}+\frac{721}{720}\right)\int_{0}^{1}w(x)dx
 +\left(-\frac{x^{3}}{6}+\frac{x^{2}}{4}-\frac{x}{12}\right)\int_{0}^{1}\int_{0}^{x}w(y)dydx\\[3pt]
&\quad +\left(-\frac{x^{2}}{2}+\frac{x}{2}-\frac{1}{12}\right)\int_{0}^{1}\int_{0}^{x}\int_{0}^{y}w(z)dzdydx
+\left(-x+\frac{1}{2}\right)\int_{0}^{1}\int_{0}^{x}\int_{0}^{y}\int_{0}^{z}w(r)drdzdydx\\[3pt]
&\quad
-\int_{0}^{1}\int_{0}^{x}\int_{0}^{y}\int_{0}^{z}\int_{0}^{r}w(s)dsdrdzdydx
+\int_{0}^{x}\int_{0}^{y}\int_{0}^{z}\int_{0}^{r}w(s)dsdrdzdy.
\end{array}
\eqno(2.1)
$$
Since the following identities hold
$$
\begin{array}{rl}
\Lambda_{\mu}^{-4}\partial_{x}w
&=\left(-\frac{x^{3}}{6}+\frac{x^{2}}{4}-\frac{x}{12}\right)\int_{0}^{1}w(x)dx
 +\left(-\frac{x^{2}}{2}+\frac{x}{2}-\frac{1}{12}\right)\int_{0}^{1}\int_{0}^{x}w(y)dydx\\[3pt]
&\quad +\left(-x+\frac{1}{2}\right)\int_{0}^{1}\int_{0}^{x}\int_{0}^{y}w(z)dzdydx
-\int_{0}^{1}\int_{0}^{x}\int_{0}^{y}\int_{0}^{z}w(r)drdzdydx\\[3pt]
&\quad
+\int_{0}^{x}\int_{0}^{z}\int_{0}^{r}w(s)dsdrdz\\[3pt]
&=\partial_{x}\Lambda_{\mu}^{-4}w,
\end{array}
\eqno(2.2)
$$
we know $\Lambda_{\mu}^{-4}$ and $\partial_{x}$ commute.

For any $s\in \mathbb{R}$, $H^{s}(\mathbb{S})$ is defined by the Sobolev space of periodic functions
$$
H^{s}(\mathbb{S})
=\{v=\sum_{k}\widehat{v}(k)e^{2\pi ikx}:~
\|v\|_{H^{s}(\mathbb{S})}^{2}=\sum |\widehat{\Lambda^{s}v}(k)|^{2}<\infty\},
$$
where the pseudodifferential operator $\Lambda^{s}=(1-\partial_{x}^{2})^{\frac{s}{2}}$ is defined by
$$
\widehat{\Lambda^{s}v}(k)=(1+4\pi^{2}k^{2})^{\frac{s}{2}}\widehat{v}(k).
$$
We can check that $\Lambda_{\mu}^{4}=(\mu-\partial_{x}^{2})^{2}$ is an isomorphism between $H^{s}(\mathbb{S})$ and $H^{s-4}(\mathbb{S})$.
Moreover, when $w\in H^{r+j-4}(\mathbb{S)}$ for $j=1,2,3$, we have
$\Lambda_{\mu}^{-4}\partial_{x}^{j}w\in H^{r}(\mathbb{S)}$ with
$$
\begin{array}{rl}
\|\Lambda_{\mu}^{-4}\partial_{x}^{j}w\|_{H^{r}(\mathbb{S)}}^{2}
&=\sum_{k}(1+4\pi^{2}k^{2})^{r}|\widehat{\Lambda_{\mu}^{-4}\partial_{x}^{j}w}(k)|^{2}\\[3pt]
&=\sum_{k}(1+4\pi^{2}k^{2})^{r}|\frac{(2\pi ik)^{j}}{\delta_{0}(k)+(2\pi k)^{4}}\widehat{w}(k)|^{2}\\[3pt]
&=\sum_{k}(1+4\pi^{2}k^{2})^{r+j-4}(1+4\pi^{2}k^{2})^{4-j}|\frac{(2\pi ik)^{j}}{\delta_{0}(k)+(2\pi k)^{4}}|^{2}|\widehat{w}(k)|^{2}\\[3pt]
&\leq 2^{4-j}\sum_{k}(1+4\pi^{2}k^{2})^{r+j-4}|\widehat{w}(k)|^{2}\\[3pt]
&=2^{4-j}\|w\|_{H^{r+j-4}(\mathbb{S)}}^{2}.
\end{array}
\eqno(2.3)
$$

\subsection{Local Well-posedness}

The initial-value problem associated to (1.1) can be rewritten in the following form:
$$
\begin{array}{l}
\left\{\begin{array}{l}
\mu(u_{t})+u_{txxxx}+2\mu(u)u_{x}+2u_{x}u_{xxxx}+uu_{xxxxx}=0, \quad t>0, ~x\in \mathbb{R}, \\[3pt]
u(t,x+1)=u(t,x), \quad t\geq0, ~x\in \mathbb{R},\\[3pt]
u(0,x)=u_{0}(x), \quad x\in \mathbb{R}.
\end{array}
\right.
\end{array}
\eqno(2.4)
$$
or, equivalently,
$$
\begin{array}{l}
\left\{\begin{array}{l}
u_{t}+uu_{x}
 +\partial_{x}\Lambda_{\mu}^{-4}\left(2\mu(u)u-3u_{x}u_{xxx}-\frac{7}{2}u_{xx}^{2}\right)=0,\quad t>0, ~x\in \mathbb{R},\\[3pt]
u(t,x+1)=u(t,x), \quad t\geq0, ~x\in \mathbb{R},\\[3pt]
u(0,x)=u_{0}(x), \quad x\in \mathbb{R},
\end{array}
\right.
\end{array}
\eqno(2.5)
$$

On the other hand, integrating both sides of (2.4) over $\mathbb{S}$ with respect to $x$, we obtain
$$
\frac{d}{dt}\mu(u)=0.
$$
Then it follows that
$$
\mu(u)=\mu(u_{0}):=\mu_{0}.
$$
Thus, (2.5) can be rewritten as
$$
\begin{array}{l}
\left\{\begin{array}{l}
u_{t}+uu_{x}
 +\partial_{x}\Lambda_{\mu}^{-4}\left(2\mu_{0}u
 -3u_{x}u_{xxx}-\frac{7}{2}u_{xx}^{2}\right)=0, \quad t>0, ~x\in \mathbb{S},\\[3pt]
u(0,x)=u_{0}(x), \quad x\in \mathbb{S}.
\end{array}
\right.
\end{array}
\eqno(2.6)
$$

Applying the Kato's theorem in \cite{k75}, one may follow the similar argument as in \cite{ly14} to obtain the following local well-posedness result for (2.6).\\

\noindent\textbf{Theorem 2.1.}  Given $u_{0}\in H^{s}(\mathbb{S}), s>\frac{7}{2}$,
there exist a maximal $T=T(u_{0})>0$, and a unique solution
$u$ to (2.6) such that
$$
u=u(\cdot,u_{0})\in C([0,T);H^{s}(\mathbb{S}))\cap C^{1}([0,T);H^{s-1}(\mathbb{S})).
$$
Moreover, the solution depends continuously on the initial data, and $T$ is independent of $s$.\\

\noindent\textbf{Lemma 2.1.}
(See \cite{ca00})
Assume $f(x)\in H^{1}(\mathbb{S})$ is such that $\int_{\mathbb{S}}f(x)dx=\frac{a_{0}}{2}$.
Then, for any $\varepsilon>0$, we have
$$
\max_{x\in\mathbb{S}}f^{2}(x)
\leq \frac{\varepsilon+2}{24}\int_{\mathbb{S}}f_{x}^{2}(x)dx+\frac{\varepsilon+2}{4\varepsilon}a_{0}^{2}.
$$

\noindent\textbf{Corollary 2.1.}
For $f(x)\in H^{1}(\mathbb{S})$, if $\int_{\mathbb{S}}f(x)dx=0$,
then we have
$$
\max_{x\in\mathbb{S}}f^{2}(x)
\leq \frac{1}{12}\int_{\mathbb{S}}f_{x}^{2}(x)dx.
$$

\noindent\textbf{Lemma 2.2.}
Let $u_{0}\in H^{s}(\mathbb{S}), s>\frac{7}{2}$,
and let $T$ be the maximal existence time of the solution $u$ to (2.6) with the initial data $u_{0}$.
Then we have
$$
\begin{array}{l}
\int_{\mathbb{S}}u_{xx}^{2}dx
    =\int_{\mathbb{S}}u_{0,xx}^{2}dx:=\mu_{1}^{2}, \quad \forall ~t\in [0,T).
\end{array}
\eqno(2.7)
$$
Moreover, we have
$$
\|u_{x}(t,\cdot)\|_{L^{\infty}(\mathbb{S})}
\leq \frac{\sqrt{3}}{6}\mu_{1}
$$
and
$$
\|u(t,\cdot)\|_{L^{\infty}(\mathbb{S})}
\leq |\mu_{0}|+\frac{1}{12}\mu_{1}.
$$

\noindent\textbf{Proof.}
A direct computation gives
$$
\frac{d}{dt}\int_{\mathbb{S}}u_{xx}^{2}dx=0,
\quad\mbox{which implies (2.7)}.
$$
Since $u(t, \cdot)\in H^{s}(\mathbb{S})\subset C^{2}(\mathbb{S})$ for $s>\frac{7}{2}$, and $\int_{\mathbb{S}}u_{x}dx=0$, Corollary 2.1 and $(2.7)$ implies that
$$
\max_{x\in\mathbb{S}}u_{x}^{2}(t,x)
\leq \frac{1}{12}\int_{\mathbb{S}}u_{xx}^{2}(t,x)dx
=\frac{1}{12}\int_{\mathbb{S}}u_{0,xx}^{2}dx
=\frac{1}{12}\mu_{1}^{2}.
$$
It then follows that
$$
\|u_{x}(t,\cdot)\|_{L^{\infty}(\mathbb{S})}
\leq \frac{\sqrt{3}}{6}\mu_{1}.
$$
Note that
$$
\int_{\mathbb{S}}(u(t,x)-\mu_{0})=0.
$$
By Corollary 2.1, we have
$$
\max_{x\in\mathbb{S}}(u(t,x)-\mu_{0})^{2}
\leq \frac{1}{12}\int_{\mathbb{S}}u_{x}^{2}(t,x)dx
\leq \frac{1}{12}\|u_{x}(t,\cdot)\|_{L^{\infty}(\mathbb{S})}^{2},
$$
which implies that
$$
\|u(t,\cdot)\|_{L^{\infty}(\mathbb{S})}-|\mu_{0}|
\leq \|u(t,\cdot)-\mu_{0}\|_{L^{\infty}(\mathbb{S})}
\leq \frac{1}{12}\mu_{1}.
$$
Hence, we get
$$
\|u(t,\cdot)\|_{L^{\infty}(\mathbb{S})}
\leq |\mu_{0}|+\frac{1}{12}\mu_{1}.
$$
This completes the proof of the lemma.  \hfill $\Box$

\section{Global Existence of Strong Solution}

In this section, we present the global existence of strong solution to (2.6). Firstly, we will give some useful lemmas.\\

\noindent\textbf{Lemma 3.1.} (see \cite{kap88}) If $r>0$, then $H^{r}(\mathbb{S})\cap L^{\infty}(\mathbb{S})$
is an algebra. Moreover,
$$
\|fg\|_{H^{r}(\mathbb{S})}
\leq c_{r}(\|f\|_{L^{\infty}(\mathbb{S})}\|g\|_{H^{r}(\mathbb{S})}
 +\|f\|_{H^{r}(\mathbb{S})}\|g\|_{L^{\infty}(\mathbb{S})}),
$$
where $c_{r}$ is a positive constant depending only on $r$.\\

\noindent\textbf{Lemma 3.2.} (see \cite{kap88}) If $r>0$, then
$$
\left\|[\Lambda^{r}, f]g\right\|_{L^{2}(\mathbb{S})}
\leq c_{r}(\|\partial_{x}f\|_{L^{\infty}(\mathbb{S})}\|\Lambda^{r-1}g\|_{L^{2}(\mathbb{S})}
 +\|\Lambda^{r}f\|_{L^{2}(\mathbb{S})}\|g\|_{L^{\infty}(\mathbb{S})}),
$$
where $\Lambda^{r}=(1-\partial_{x}^{2})^{r/2}$ and $c_{r}$ is a positive constant depending only on $r$.\\

\noindent\textbf{Lemma 3.3.} (see \cite{esko14,tay14}) If $f\in H^{s}(\mathbb{S})$ with $s>\frac{3}{2}$, then there exists a constant $c>0$ such that for any $g\in L^{2}(\mathbb{S})$ we have
$$
\|[J_{\varepsilon}, f]\partial_{x}g\|_{L^{2}(\mathbb{S})}
\leq c\|f\|_{C^{1}(\mathbb{S})}\|g\|_{L^{2}(\mathbb{S})},
$$
in which for each $\varepsilon\in(0,1]$, the operator $J_{\varepsilon}$ is the Friedrichs mollifier defined by
$$
\begin{array}{l}
J_{\varepsilon}f(x)=j_{\varepsilon}*f(x),
\end{array}
\eqno(3.1)
$$
where $j_{\varepsilon}(x)=\frac{1}{\varepsilon}j(\frac{x}{\varepsilon})$ and $j(x)$ is a nonnegative, even, smooth bump function supported in the interval $(-\frac{1}{2}, \frac{1}{2})$ such that $\int_{\mathbb{R}}j(x)dx=1$.
For any $f\in H^{s}(\mathbb{S})$ with $s\geq0$, we have $J_{\varepsilon}f\rightarrow f$ in $H^{s}(\mathbb{S})$ as $\varepsilon\rightarrow 0$. Moreover, for any $p\geq1$, the Young's inequality
$\|j_{\varepsilon}*f\|_{L^{p}(\mathbb{S})}\leq \|j_{\varepsilon}\|_{L^{1}(\mathbb{R})}\|f\|_{L^{p}(\mathbb{S})}=\|f\|_{L^{p}(\mathbb{S})}$ holds since $j(x)$ is supported in the interval $(-\frac{1}{2}, \frac{1}{2})$.\\

Now we give the following theorem, which is a sufficient condition of global existence of solution to (2.6).\\

\noindent\textbf{Theorem 3.1.} Let $u_{0}\in H^{s}(\mathbb{S}), s>\frac{7}{2}$,
and let $T$ be the maximal existence time of the solution $u$ to (2.6) with the initial data $u_{0}$. If there exists $K>0$ such that
$$
\|u_{xxx}(t,\cdot)\|_{L^{\infty}(\mathbb{S})}\leq K, \quad t\in [0,T),
$$
then the $H^{s}(\mathbb{S})$-norm of $u(t,\cdot)$ does not blow up on $[0,T)$.\\

\noindent\textbf{Proof.}
Note that the product $uu_{x}$ only has the regularity of $H^{s-1}(\mathbb{S})$ when $u\in H^{s}(\mathbb{S})$. To deal with this problem, we will consider the following modified equation
$$
\begin{array}{l}
(J_{\varepsilon}u)_{t}+J_{\varepsilon}(uu_{x})
 +\partial_{x}\Lambda_{\mu}^{-4}\left(2\mu_{0}J_{\varepsilon}u
 -3J_{\varepsilon}(u_{x}u_{xxx})-\frac{7}{2}J_{\varepsilon}(u_{xx}^{2})\right)=0,
\end{array}
\eqno(3.2)
$$
where $J_{\varepsilon}$ is defined in (3.1).

Applying the operator $\Lambda^{s}=(1-\partial_{x}^{2})^{s/2}$ to (3.2), then multiplying the resulting equation
by $\Lambda^{s}J_{\varepsilon}u$ and integrating with respect to $x\in \mathbb{S}$, we obtain
$$
\begin{array}{rl}
\frac{1}{2}\frac{d}{dt}\|J_{\varepsilon}u\|_{H^{s}(\mathbb{S})}^{2}
&=-\left(\Lambda^{s}J_{\varepsilon}(uu_{x}), \Lambda^{s}J_{\varepsilon}u\right)\\[3pt]
&\quad -\left(\Lambda^{s}J_{\varepsilon}u,\partial_{x}\Lambda^{s}\Lambda_{\mu}^{-4}\left(2\mu_{0}J_{\varepsilon}u
     -3J_{\varepsilon}(u_{x}u_{xxx})-\frac{7}{2}J_{\varepsilon}(u_{xx}^{2})\right)\right).
\end{array}
\eqno(3.3)
$$
In what follows next we use the fact that $\Lambda^{s}$ and $J_{\varepsilon}$ commute and that
$J_{\varepsilon}$ satisfies the properties
$$
(J_{\varepsilon}f, g)=(f, J_{\varepsilon}g)
\quad \mbox{and}\quad
\|J_{\varepsilon}u\|_{H^{s}(\mathbb{S})}\leq \|u\|_{H^{s}(\mathbb{S})}.
$$
Let us estimate the first term of the right hand side of (3.3).
$$
\begin{array}{rl}
&\left|\left(\Lambda^{s}J_{\varepsilon}(uu_{x}), \Lambda^{s}J_{\varepsilon}u\right)\right|\\[3pt]
&=\left|(\Lambda^{s}(uu_{x}),J_{\varepsilon}\Lambda^{s}J_{\varepsilon}u)\right|\\[3pt]
&=\left|([\Lambda^{s}, u]u_{x},J_{\varepsilon}\Lambda^{s}J_{\varepsilon}u)
  +( u\Lambda^{s}u_{x},J_{\varepsilon}\Lambda^{s}J_{\varepsilon}u)\right|\\[3pt]
&=\left|([\Lambda^{s}, u]u_{x},J_{\varepsilon}\Lambda^{s}J_{\varepsilon}u)
  +( J_{\varepsilon}u\partial_{x}\Lambda^{s}u,\Lambda^{s}J_{\varepsilon}u)\right|\\[3pt]
&=\left|([\Lambda^{s}, u]u_{x},J_{\varepsilon}\Lambda^{s}J_{\varepsilon}u)
  +([J_{\varepsilon},u]\partial_{x}\Lambda^{s}u,\Lambda^{s}J_{\varepsilon}u)
  +(uJ_{\varepsilon}\partial_{x}\Lambda^{s}u,\Lambda^{s}J_{\varepsilon}u)\right|\\[3pt]
&\leq\|[\Lambda^{s}, u]u_{x}\|_{L^{2}(\mathbb{S})}
  \|J_{\varepsilon}\Lambda^{s}J_{\varepsilon}u\|_{L^{2}(\mathbb{S})}
  +\|[J_{\varepsilon},u]\partial_{x}\Lambda^{s}u\|_{L^{2}(\mathbb{S})}
  \|\Lambda^{s}J_{\varepsilon}u\|_{L^{2}(\mathbb{S})}\\[3pt]
&\quad+\frac{1}{2}\left|(u_{x}\Lambda^{s}J_{\varepsilon}u,\Lambda^{s}J_{\varepsilon}u)\right|\\[3pt]
&\lesssim \|u\|_{C^{1}(\mathbb{S})}\|u\|_{H^{s}(\mathbb{S})}^{2},
\end{array}
\eqno(3.4)
$$
where we have used Lemma 3.2 with $r=s$ and Lemma 3.3. Here and in what follows, we use $``\lesssim"$ to denote inequality up to a positive constant. Furthermore, we estimate the second term of the right hand side of $(3.3)$ in the following way
$$
\begin{array}{rl}
&\left|\left(\Lambda^{s}J_{\varepsilon}u,\partial_{x}\Lambda^{s}\Lambda_{\mu}^{-4}\left(2\mu_{0}J_{\varepsilon}u
     -3J_{\varepsilon}(u_{x}u_{xxx})-\frac{7}{2}J_{\varepsilon}(u_{xx}^{2})\right)\right)\right|\\[3pt]
&\leq\left\|\partial_{x}\Lambda_{\mu}^{-4}\left(2\mu_{0}J_{\varepsilon}u
     -3J_{\varepsilon}(u_{x}u_{xxx})-\frac{7}{2}J_{\varepsilon}(u_{xx}^{2})\right)\right\|_{H^{s}(\mathbb{S})}\|u\|_{H^{s}(\mathbb{S})}\\[3pt]
&\lesssim\|2\mu_{0}J_{\varepsilon}u
     -3J_{\varepsilon}(u_{x}u_{xxx})-\frac{7}{2}J_{\varepsilon}(u_{xx}^{2})\|_{H^{s-3}(\mathbb{S})}\|u\|_{H^{s}(\mathbb{S})}\\[3pt]
&\lesssim (|\mu_{0}|\|u\|_{H^{s-3}(\mathbb{S})}
  +\|u_{x}\|_{L^{\infty}(\mathbb{S})}\|u\|_{H^{s}(\mathbb{S})}
  +\|u_{xxx}\|_{L^{\infty}(\mathbb{S})}\|u\|_{H^{s-2}(\mathbb{S})}\\[3pt]
&\qquad  +\|u_{xx}\|_{L^{\infty}(\mathbb{S})}\|u_{xx}\|_{H^{s-3}(\mathbb{S})})\|u\|_{H^{s}(\mathbb{S})}\\[3pt]
&\leq \left(|\mu_{0}|
  +\|u_{x}\|_{L^{\infty}(\mathbb{S})}
  +\|u_{xx}\|_{L^{\infty}(\mathbb{S})}
  +\|u_{xxx}\|_{L^{\infty}(\mathbb{S})}\right)\|u\|_{H^{s}(\mathbb{S})}^{2},
\end{array}
$$
where we have used Lemma 3.1 with $r=s-3$ and (2.3).

Since $u(t, \cdot)\in H^{s}(\mathbb{S})\subset C^{2}(\mathbb{S})$ for $s>\frac{7}{2}$, and $\int_{\mathbb{S}}u_{xx}dx=0$, Corollary 2.1 implies that
$$
\|u_{xx}(t,\cdot)\|_{L^{\infty}(\mathbb{S})}
\leq \frac{\sqrt{3}}{6}\|u_{xxx}(t,\cdot)\|_{L^{2}(\mathbb{S})}
\leq \frac{\sqrt{3}}{6}\|u_{xxx}(t,\cdot)\|_{L^{\infty}(\mathbb{S})}, \quad t\in [0,T).
$$
Thus,
$$
\begin{array}{rl}
&\left|\left(\Lambda^{s}J_{\varepsilon}u,\partial_{x}\Lambda^{s}\Lambda_{\mu}^{-4}\left(2\mu_{0}J_{\varepsilon}u
     -3J_{\varepsilon}(u_{x}u_{xxx})-\frac{7}{2}J_{\varepsilon}(u_{xx}^{2})\right)\right)\right|\\[3pt]
&\lesssim \left(|\mu_{0}|
  +\|u_{x}\|_{L^{\infty}(\mathbb{S})}
  +\|u_{xxx}\|_{L^{\infty}(\mathbb{S})}\right)\|u\|_{H^{s}(\mathbb{S})}^{2},
\end{array}
\eqno(3.5)
$$
Combining $(3.4)$ and $(3.5)$
and using Lemma 2.2, we have
$$
\begin{array}{rl}
\frac{1}{2}\frac{d}{dt}\|J_{\varepsilon}u\|_{H^{s}(\mathbb{S})}^{2}
&\lesssim \left(|\mu_{0}|
  +\|u\|_{L^{\infty}(\mathbb{S})}
  +\|u_{x}\|_{L^{\infty}(\mathbb{S})}
  +\|u_{xxx}\|_{L^{\infty}(\mathbb{S})}\right)\|u\|_{H^{s}(\mathbb{S})}^{2}\\[5pt]
&\lesssim (|\mu_{0}|+\mu_{1}+\|u_{xxx}\|_{L^{\infty}(\mathbb{S})})\|u\|_{H^{s}(\mathbb{S})}^{2}.
\end{array}
\eqno(3.6)
$$
Letting $\varepsilon\rightarrow 0$, we get
$$
\begin{array}{l}
\frac{1}{2}\frac{d}{dt}\|u\|_{H^{s}(\mathbb{S})}^{2}
\leq c(1+\|u_{xxx}\|_{L^{\infty}(\mathbb{S})})\|u\|_{H^{s}(\mathbb{S})}^{2},
\end{array}
$$
where $c$ is a constant depending on $s$ and $u_{0}$.
An application of Gronwall's inequality and the assumption of the theorem yield
$$
\|u\|_{H^{s}(\mathbb{S})}^{2}
\leq e^{2c(1+K)t}\|u_{0}\|_{H^{s}(\mathbb{S})}^{2},
$$
which completes the proof of the theorem.  \hfill $\Box$\\

\noindent\textbf{Theorem 3.2.}
Let $u_{0}\in H^{s}(\mathbb{S}), s>\frac{7}{2}$. Then the corresponding strong solution $u$ of the initial value $u_{0}$ exists globally in time.\\

\noindent\noindent\textbf{Proof.} Let us first assume that $u_{0}\in H^{s}(\mathbb{S})$, $s\geq 5$.  Let $u$ be the corresponding solution of (2.6) on $[0, T)\times \mathbb{S}$, which is guaranteed by Theorem 2.1. Multiplying (1.1) by $m$ and integrating over $\mathbb{S}$ with respect to $x$ yield
$$
\frac{1}{2}\frac{d}{dt}\int_{\mathbb{S}}m^{2}dx
=\int_{\mathbb{S}}m(-2mu_{x}-um_{x})dx
=-3\int_{\mathbb{S}}u_{x}m^{2}dx.
\eqno(3.7)
$$
Note that in Lemma 2.2 we have
$$
\|u_{x}(t,\cdot)\|_{L^{\infty}(\mathbb{S})}
\leq \frac{\sqrt{3}}{6}\mu_{1}.
$$
Then
$$
\frac{d}{dt}\int_{\mathbb{S}}m^{2}dx
\leq \sqrt{3}\mu_{1}\int_{\mathbb{S}}m^{2}dx.
$$
By Gronwall's inequality, we have
$$
\int_{\mathbb{S}}m^{2}dx\leq e^{\sqrt{3}\mu_{1}t}\int_{\mathbb{S}}m_{0}^{2}dx.
$$
Note that
$$
\int_{\mathbb{S}}m^{2}
=\mu(u)^{2}+\int_{\mathbb{S}}u_{xxxx}^{2}\geq\|u_{xxxx}\|_{L^{2}(\mathbb{S})}^{2}.
$$
Since $u\in H^{5}(\mathbb{S})\subset C^{4}(\mathbb{S})$ and $\int_{\mathbb{S}}u_{xxx}dx=0$,
Corollary 2.1 implies that
$$
\begin{array}{l}
\|u_{xxx}\|_{L^{\infty}(\mathbb{S})}
\leq \frac{\sqrt{3}}{6}\|u_{xxxx}\|_{L^{2}(\mathbb{S})}
\leq \frac{\sqrt{3}}{6}\|m\|_{L^{2}(\mathbb{S})}
\leq \frac{\sqrt{3}}{6}e^{\frac{\sqrt{3}}{2}\mu_{1}t}\|m_{0}\|_{L^{2}(\mathbb{S})}.
\end{array}
$$
Theorem 3.1 ensures that the solution $u$ does not blow up in finite time, that is, $T=\infty$.

For initial data $u_{0}\in H^{s}(\mathbb{S}), \frac{7}{2}<s<5$, we can approximate $u_{0}$ in $H^{s}(\mathbb{S})$ by the functions $u_{0}^{n}\in H^{5}(\mathbb{S})$. By Theorem 2.1, there
exist the corresponding solutions
$$
u^{n}\in C([0, T_{n}); H^{5}(\mathbb{S}))\cap C^{1}([0, T_{n}); H^{4}(\mathbb{S}))
$$
with initial data $u_{0}^{n}$, $n\geq 1$. From the above analysis, we know $T_{n}=\infty$.
Since by Theorem 2.1 the maximal time continuously depends on $\|u_0\|_{H^{s}(\mathbb{S})}$ and $u_{0}^{n}\rightarrow u_{0}$ in $H^{s}(\mathbb{S})$, it follows that $T_{n}\rightarrow T$ as $n\rightarrow\infty$. Thus, we obtain $T=\infty$. This completes the proof
of Theorem 3.2. \hfill $\Box$

\section{H\"{o}lder continuity}

In this section, we will first give an estimate of the solution size in time interval
$[0, T_{0}]$ for any fixed $T_{0}>0$,
and then we show that the solution map for (2.6) is H\"{o}lder continuous
in $H^{s}(\mathbb{S})$, $s\geq 4$, equipped with the $H^{r}(\mathbb{S})$-topology for $0\leq r<s$ in periodic case.\\

\noindent\textbf{Lemma 4.1.} Let $u$ be the solution of (2.6) with initial data $u_{0}\in H^{s}(\mathbb{S}), s\geq 4$. Then, for any fixed $T_{0}>0$, we have
$$
\|u(t)\|_{H^{s}(\mathbb{S})}\leq
e^{cT_{0}}\|u_{0}\|_{H^{s}(\mathbb{S})},
\quad t\in[0, T_{0}],
$$
where $c=c(s, T_{0}, u_{0})$ is a constant depending on $s$, $T_{0}$ and $\|u_{0}\|_{H^{4}(\mathbb{S})}$.\\

\noindent\textbf{Proof.}
By using the local well-posedness theorem and a density argument, we can show $(3.7)$ still holds
for initial data $u_{0}\in H^{s}(\mathbb{S})$, $s\geq 4$, that is,
$$
\frac{1}{2}\frac{d}{dt}\int_{\mathbb{S}}m^{2}dx
=-3\int_{\mathbb{S}}u_{x}m^{2}dx.
$$
Then, similar to the proof of Theorem 3.2, we can show that
$$
\begin{array}{l}
\|u_{xxx}\|_{L^{\infty}(\mathbb{S})}
\leq \frac{\sqrt{3}}{6}\|u_{xxxx}\|_{L^{2}(\mathbb{S})}
\leq \frac{\sqrt{3}}{6}\|m\|_{L^{2}(\mathbb{S})}
\leq \frac{\sqrt{3}}{6}e^{\frac{\sqrt{3}}{2}\mu_{1}t}\|m_{0}\|_{L^{2}(\mathbb{S})}
\leq \frac{\sqrt{3}}{6}e^{\frac{\sqrt{3}}{2}\mu_{1}T_{0}}\|m_{0}\|_{L^{2}(\mathbb{S})}.
\end{array}
$$
Note that in $(3.6)$ we have
$$
\begin{array}{rl}
\frac{1}{2}\frac{d}{dt}\|u\|_{H^{s}(\mathbb{S})}^{2}
&\lesssim \left(|\mu_{0}|
  +\|u\|_{L^{\infty}(\mathbb{S})}
  +\|u_{x}\|_{L^{\infty}(\mathbb{S})}
  +\|u_{xxx}\|_{L^{\infty}(\mathbb{S})}\right)\|u\|_{H^{s}(\mathbb{S})}^{2}\\[3pt]
&\lesssim (|\mu_{0}|+\mu_{1}+\|u_{xxx}\|_{L^{\infty}(\mathbb{S})})\|u\|_{H^{s}(\mathbb{S})}^{2}\\[3pt]
&\lesssim (|\mu_{0}|+\mu_{1}+e^{\frac{\sqrt{3}}{2}\mu_{1}T_{0}}\|m_{0}\|_{L^{2}(\mathbb{S})})\|u\|_{H^{s}(\mathbb{S})}^{2}, \quad\forall~0\leq t\leq T_{0}.
\end{array}
$$
It follows that
$$
\|u\|_{H^{s}(\mathbb{S})}\leq e^{cT_{0}}\|u_{0}\|_{H^{s}(\mathbb{S})},\quad\forall~0\leq t\leq T_{0},
$$
where $c=c(s, T_{0}, u_{0})$ is a constant depending on $s$, $T_{0}$ and $\|u_{0}\|_{H^{4}(\mathbb{S})}$. \hfill $\Box$\\

Now, we prove that the solution map for (2.6) is H\"{o}lder continuous. Firstly, we recall the following lemma.\\

\noindent\textbf{Lemma 4.2.} (see \cite{taylor03}) If $s>\frac{3}{2}$ and $0\leq \sigma+1\leq s$, then there exists a constant $c>0$ such that
$$
\|[\Lambda^{\sigma}\partial_{x}, f]v\|_{L^{2}(\mathbb{S})}
\leq c\|f\|_{H^{s}(\mathbb{S})}\|v\|_{H^{\sigma}(\mathbb{S})}.
$$

\noindent\textbf{Lemma 4.3.} (see \cite{himkm10}) If $r>\frac{1}{2}$, then there exists a constant $c_{r}>0$ depending only on $r$ such that
$$
\|fg\|_{H^{r-1}(\mathbb{S})}\leq c_{r}\|f\|_{H^{r}(\mathbb{S})}\|g\|_{H^{r-1}(\mathbb{S})}.
$$

Lemma 4.3 gives the estimate of $\|fg\|_{H^{s}(\mathbb{S})}$ for $s>-\frac{1}{2}$, the other cases are provided in the following lemma.\\

\noindent\textbf{Lemma 4.4.} If $0\leq r\leq j$, $l>\frac{1}{2}$ and $l\geq j-r$ with $j\in \mathbb{Z}_{+}$, then
there exists a constant $c_{r,l,j}>0$ depending on $r$, $l$ and $j$ such that
$$
\|fg\|_{H^{r-j}(\mathbb{S})}\leq c_{r,l,j}\|f\|_{H^{l}(\mathbb{S})}\|g\|_{H^{r-j}(\mathbb{S})}.
$$

\noindent\textbf{Proof.} The proof can be done by adapting analogous methods as in \cite{himho13}, in which they only considered the case $j=1$. For the reader's convenience, we provide the arguments with obvious modifications. Similar as the proof of Lemma 3 on $\mathbb{R}$ in \cite{himho13}, we can obtain
$$
\begin{array}{rl}
\|fg\|_{H^{r-j}(\mathbb{S})}^{2}
&=\sum_{k}(1+k^{2})^{r-j}|\sum_{n}\widehat{f}(n)\widehat{g}(k-n)|^{2}\\[3pt]
&=\sum_{k}(1+k^{2})^{r-j}|\sum_{n}(1+n^{2})^{\frac{l}{2}}\widehat{f}(n)\cdot(1+n^{2})^{-\frac{l}{2}}\widehat{g}(k-n)|^{2}\\[3pt]
&\leq\|f\|_{H^{l}(\mathbb{S})}^{2}\sum_{n}|\widehat{g}(n)|^{2}
\sum_{k}(1+k^{2})^{r-j}(1+(k-n)^{2})^{-l},
\end{array}
$$
in which we have applied the Cauchy-Schwartz inequality in $n$, a change of variables, and changed the order of summation.
To get the desired result, it is sufficient to show that there exists a constant
$c_{r,l,j}>0$ such that
$$
\begin{array}{rl}
\sum_{k}(1+k^{2})^{r-j}(1+(k-n)^{2})^{-l}
\leq c_{r,l,j}(1+n^{2})^{r-j}.
\end{array}
$$
In fact, we can check the inequality under the conditions $l>\frac{1}{2}$ and $l\geq j-r$,
the main difference with proof of Lemma 5 in \cite{himho13} is replacing the discussions on $\frac{1}{2}<r\leq 1$ ($0\leq r< \frac{1}{2}$, $r=\frac{1}{2}$, respectively) by $j-\frac{1}{2}<r\leq j$ ($0\leq r< j-\frac{1}{2}$, $r=j-\frac{1}{2}$, respectively) for the case $k\in [\lceil \frac{n}{2}\rceil, n]$, where
$\lceil \frac{n}{2}\rceil=[\frac{n}{2}]+1$ and
$[\frac{n}{2}]$ is the integer part of $\frac{n}{2}$.\hfill $\Box$\\

\noindent\textbf{Remark 4.1.} Lemma 4.4 is more general than $(iii)$ of Proposition 2.4 in \cite{guil10} when considering the Sobolev norm of $fg$ with negative index, since it covers the case $l=j-r$ here.\\

\noindent\textbf{Theorem 4.1.} Assume $s\geq4$ and $0\leq r<s$. Then the solution map
for (2.6) is H\"{o}lder continuous with exponent
$$
\begin{array}{rl}
\alpha=
\left\{\begin{array}{l}
1, \quad \mbox{if}~0\leq r\leq s-1, \\[3pt]
s-r, \quad \mbox{if}~s-1< r< s
\end{array}
\right.
\end{array}
$$
as a map from $B(0,h):=\{u\in H^{s}(\mathbb{S}): \|u\|_{H^{s}(\mathbb{S})}\leq h\}$ with $H^{r}(\mathbb{S})$-norm to $C([0,T_{0}]; H^{r}(\mathbb{S}))$
for any fixed $T_{0}>0$. More precisely, we have
$$
\|u(t)-w(t)\|_{C([0,T_{0}]; H^{r}(\mathbb{S}))}
\leq c\|u(0)-w(0)\|_{H^{r}(\mathbb{S})}^{\alpha},
$$
for all $u(0), w(0)\in B(0,h)$
and $u(t), w(t)$ the solutions corresponding to the initial data $u(0), w(0)$,
respectively. The constant $c$ depends on $s, r, T_{0}$ and $h$.\\

\noindent\textbf{Proof.} Define $v=u-w$, then $v$ satisfies that
$$
\begin{array}{rl}
\left\{\begin{array}{l}
v_{t}+\partial_{x}\left(\frac{1}{2}v(u+w)\right)\\[3pt]
\quad+\partial_{x}\Lambda_{\mu}^{-4}\left(2\mu(u)v+2\mu(v)w-3w_{x}v_{xxx}-3v_{x}u_{xxx}-\frac{7}{2}v_{xx}(u_{xx}+w_{xx})\right)=0,\\[3pt]
v(0,x)=v_{0}(x), \quad x\in \mathbb{S}.
\end{array}
\right.
\end{array}
\eqno(4.1)
$$
Applying $\Lambda^{r}$ to both sides of (4.1), then multiplying both sides by $\Lambda^{r}v$ and integrating over $\mathbb{S}$ with respect to $x$, we get
$$
\begin{array}{rl}
&\frac{1}{2}\frac{d}{dt}\|v\|_{H^{r}(\mathbb{S})}^{2}\\[3pt]
&=-\int_{\mathbb{S}}\Lambda^{r}\partial_{x}
  \left(\frac{1}{2}v(u+w)\right)\cdot \Lambda^{r}vdx\\[3pt]
&\quad-\int_{\mathbb{S}}\Lambda^{r}\partial_{x}\Lambda_{\mu}^{-4}\left(2\mu(u)v+2\mu(v)w-3w_{x}v_{xxx}-3v_{x}u_{xxx}-\frac{7}{2}v_{xx}(u_{xx}+w_{xx})\right)\cdot \Lambda^{r}vdx\\[3pt]
&:=E_{1}+E_{2}.
\end{array}
$$

$(i)$ We first consider the case $0\leq r\leq s-1$, where $s\geq4$. To achieve it, we need to estimate $E_{1}$ and $E_{2}$.

\emph{Estimate} $E_{1}$. By using Lemma 4.2 and the Sobolev embedding theorem $H^{r}(\mathbb{S})\hookrightarrow L^{\infty}(\mathbb{S})$ for $r>\frac{1}{2}$, we have
$$
\begin{array}{rl}
|E_{1}|
&=\left|-\int_{\mathbb{S}}\Lambda^{r}\partial_{x}
  \left(\frac{1}{2}v(u+w)\right)\cdot \Lambda^{r}vdx\right|\\[3pt]
&=\left|-\int_{\mathbb{S}}[\Lambda^{r}\partial_{x}, \frac{1}{2}(u+w)]v\cdot \Lambda^{r}vdx
  -\int_{\mathbb{S}}\frac{1}{2}(u+w)\Lambda^{r}\partial_{x}v\cdot \Lambda^{r}vdx\right|\\[3pt]
&\lesssim\|u+w\|_{H^{s}(\mathbb{S})}\|v\|_{H^{r}(\mathbb{S})}^{2}.
\end{array}
$$

\emph{Estimate} $E_{2}$. It is easy to show that
$$
\begin{array}{rl}
&|E_{2}|\\[3pt]
&=\left|-\int_{\mathbb{S}}\Lambda^{r}\partial_{x}\Lambda_{\mu}^{-4}\left(2\mu(u)v+2\mu(v)w-3w_{x}v_{xxx}-3v_{x}u_{xxx}-\frac{7}{2}v_{xx}(u_{xx}+w_{xx})\right)\cdot \Lambda^{r}vdx\right|\\[3pt]
&\leq \left\|\partial_{x}\Lambda_{\mu}^{-4}\left(2\mu(u)v+2\mu(v)w-3w_{x}v_{xxx}-3v_{x}u_{xxx}-\frac{7}{2}v_{xx}(u_{xx}+w_{xx})\right)\right\|_{H^{r}(\mathbb{S})}\|v\|_{H^{r}(\mathbb{S})}.
\end{array}
$$
Using (2.3), we have
$$
\begin{array}{rl}
&\left\|\partial_{x}\Lambda_{\mu}^{-4}\left(2\mu(u)v+2\mu(v)w\right)\right\|_{H^{r}(\mathbb{S})}\\[5pt]
&\lesssim |\mu(u)|\|v\|_{H^{r-3}(\mathbb{S)}}
+|\mu(v)|\|w\|_{H^{r-3}(\mathbb{S)}}\\[5pt]
&\lesssim (\|u\|_{H^{s}(\mathbb{S})}+\|w\|_{H^{s}(\mathbb{S})})\|v\|_{H^{r}(\mathbb{S})},
\end{array}
$$
where we have used the following inequality
$$
\begin{array}{l}
|\mu(v)|=\left|\int_{\mathbb{S}}v(t,x)dx\right|
\leq\int_{\mathbb{S}}|v(t,x)|dx
\leq\|v\|_{L^{2}(\mathbb{S})}
\leq\|v\|_{H^{r}(\mathbb{S})}\quad \mbox{for~any}~r\geq0.
\end{array}
$$
On the other hand, integrating by parts, we have
$$
\begin{array}{rl}
&\left\|\partial_{x}\Lambda_{\mu}^{-4}\left(3w_{x}v_{xxx}+3v_{x}u_{xxx}+\frac{7}{2}v_{xx}(u_{xx}+w_{xx})\right)\right\|_{H^{r}(\mathbb{S})}\\[3pt]
&=\left\|-\frac{1}{2}\Lambda_{\mu}^{-4}\partial_{x}[(w_{x}+u_{x})v_{xxx}]
+3\Lambda_{\mu}^{-4}\partial_{x}^{2}(u_{xx}v_{x})
+\frac{1}{2}\Lambda_{\mu}^{-4}\partial_{x}^{2}(u_{x}v_{xx})
+\frac{7}{2}\Lambda_{\mu}^{-4}\partial_{x}^{2}(w_{x}v_{xx})\right\|_{H^{r}(\mathbb{S})}\\[3pt]
&\lesssim \left\|\Lambda_{\mu}^{-4}\partial_{x}[(w_{x}+u_{x})v_{xxx}]\right\|_{H^{r}(\mathbb{S)}}
+\left\|\Lambda_{\mu}^{-4}\partial_{x}^{2}(u_{xx}v_{x})\right\|_{H^{r}(\mathbb{S)}}
+\left\|\Lambda_{\mu}^{-4}\partial_{x}^{2}(u_{x}v_{xx})\right\|_{H^{r}(\mathbb{S)}}\\[3pt]
&\quad+\left\|\Lambda_{\mu}^{-4}\partial_{x}^{2}(w_{x}v_{xx})\right\|_{H^{r}(\mathbb{S)}}\\[3pt]
&:=F_{1}+F_{2}+F_{3}+F_{4}.
\end{array}
$$
For $F_{1}$, by $(2.3)$,
$$
\begin{array}{rl}
F_{1}
=\left\|\Lambda_{\mu}^{-4}\partial_{x}[(w_{x}+u_{x})v_{xxx}]\right\|_{H^{r}(\mathbb{S)}}
\lesssim \|(w_{x}+u_{x})v_{xxx}\|_{H^{r-3}(\mathbb{S)}}.
\end{array}
$$
If $r>\frac{5}{2}$, we have
$$
\|(w_{x}+u_{x})v_{xxx}\|_{H^{r-3}(\mathbb{S)}}
\lesssim\|w_{x}+u_{x}\|_{H^{r-2}(\mathbb{S})}\|v_{xxx}\|_{H^{r-3}(\mathbb{S})}
\leq (\|w\|_{H^{s}(\mathbb{S})}+\|u\|_{H^{s}(\mathbb{S})})\|v\|_{H^{r}(\mathbb{S})},
$$
by using Lemma 4.3 and the fact $r\leq s-1$.
For $0<r\leq \frac{5}{2}$, applying Lemma 4.4 with $j=3$ to the term $\|(w_{x}+u_{x})v_{xxx}\|_{H^{r-3}(\mathbb{S)}}$,
we have
$$
\begin{array}{rl}
\|(w_{x}+u_{x})v_{xxx}\|_{H^{r-3}(\mathbb{S)}}
\lesssim \|w_{x}+u_{x}\|_{H^{3}(\mathbb{S)}}\|v_{xxx}\|_{H^{r-3}(\mathbb{S)}}
\leq (\|w\|_{H^{s}(\mathbb{S})}+\|u\|_{H^{s}(\mathbb{S})})\|v\|_{H^{r}(\mathbb{S})}.
\end{array}
$$
Thus, we have
$$
F_{1}
\lesssim (\|w\|_{H^{s}(\mathbb{S})}+\|u\|_{H^{s}(\mathbb{S})})\|v\|_{H^{r}(\mathbb{S})}.
$$
For $F_{2}$, by $(2.3)$,
$$
\begin{array}{rl}
F_{2}
=\left\|\Lambda_{\mu}^{-4}\partial_{x}^{2}(u_{xx}v_{x})\right\|_{H^{r}(\mathbb{S)}}
\lesssim \|u_{xx}v_{x}\|_{H^{r-2}(\mathbb{S)}}.
\end{array}
$$
If $r>\frac{3}{2}$, we have
$$
\|u_{xx}v_{x}\|_{H^{r-2}(\mathbb{S)}}
\lesssim\|u_{xx}\|_{H^{r-1}(\mathbb{S})}\|v_{x}\|_{H^{r-2}(\mathbb{S})}
\leq \|u\|_{H^{s}(\mathbb{S})}\|v\|_{H^{r}(\mathbb{S})},
$$
by using Lemma 4.3 and the fact $r\leq s-1$.
For $0<r\leq \frac{3}{2}$, applying Lemma 4.4 with $j=2$ to the term $\|u_{xx}v_{x}\|_{H^{r-2}(\mathbb{S)}}$,
we have
$$
\begin{array}{rl}
\|u_{xx}v_{x}\|_{H^{r-2}(\mathbb{S)}}
\lesssim \|u_{xx}\|_{H^{2}(\mathbb{S)}}\|v_{x}\|_{H^{r-2}(\mathbb{S)}}
\leq \|u\|_{H^{s}(\mathbb{S})}\|v\|_{H^{r}(\mathbb{S})}.
\end{array}
$$
Thus, we have
$$
F_{2}
\lesssim \|u\|_{H^{s}(\mathbb{S})}\|v\|_{H^{r}(\mathbb{S})}.
$$
Similar as the estimate of $F_{2}$, we can deduce
$$
F_{3},~F_{4}
\lesssim \|u\|_{H^{s}(\mathbb{S})}\|v\|_{H^{r}(\mathbb{S})}.
$$

Using the solution size estimate in Lemma 4.1, we get
$$
\begin{array}{rl}
\frac{1}{2}\frac{d}{dt}\|v\|_{H^{r}(\mathbb{S})}^{2}
&\lesssim (\|u\|_{H^{s}(\mathbb{S})}+\|w\|_{H^{s}(\mathbb{S})})
\|v\|_{H^{r}(\mathbb{S})}^{2}\\[5pt]
&\leq (2e^{c_{1}T_{0}}\|u_{0}\|_{H^{s}(\mathbb{S})}+2e^{c_{2}T_{0}}\|w_{0}\|_{H^{s}(\mathbb{S})})
\|v\|_{H^{r}(\mathbb{S})}^{2}\\[5pt]
&\leq 4e^{\max\{c_{1}, c_{2}\}T_{0}}h\|v\|_{H^{r}(\mathbb{S})}^{2},
\end{array}
$$
which implies that
$$
\begin{array}{l}
\|u-w\|_{H^{r}(\mathbb{S})}\leq e^{CT_{0}}\|u_{0}-w_{0}\|_{H^{r}(\mathbb{S})},
\end{array}
\eqno(4.2)
$$
where $C$ is a constant depending on $s, r, T_{0}$ and $h$.

$(ii)$ Now we consider the case $s-1<r<s$, where $s\geq4$. Interpolating between $H^{s-1}(\mathbb{S})$ and $H^{s}(\mathbb{S})$ norms, we obtain
$$
\begin{array}{l}
\|v\|_{H^{r}(\mathbb{S})}
\leq\|v\|^{s-r}_{H^{s-1}(\mathbb{S})}\|v\|^{r+1-s}_{H^{s}(\mathbb{S})}.
\end{array}
\eqno(4.3)
$$
Since for $H^{s-1}(\mathbb{S})$ norm we can apply inequality (4.2), we have
$$
\begin{array}{l}
\|v\|_{H^{s-1}(\mathbb{S})}\leq e^{CT_{0}}\|u_{0}-w_{0}\|_{H^{s-1}(\mathbb{S})}.
\end{array}
\eqno(4.4)
$$
Also, using the solution size estimate in Lemma 4.1, we get
$$
\begin{array}{l}
\|v\|_{H^{s}(\mathbb{S})}\leq e^{cT_{0}}(\|u_{0}\|_{H^{s}(\mathbb{S})}+\|w_{0}\|_{H^{s}(\mathbb{S})})
\leq 2e^{cT_{0}}h.
\end{array}
\eqno(4.5)
$$
Combining (4.3), (4.4) and (4.5) gives
$$
\|v\|_{H^{r}(\mathbb{S})}
\lesssim \|v_{0}\|^{s-r}_{H^{s-1}(\mathbb{S})}
\leq \|v_{0}\|^{s-r}_{H^{r}(\mathbb{S})}.
$$
This completes the proof of Theorem 4.1.\hfill $\Box$

\section{Global Existence of Weak Solution}

In this section, we establish the existence of global weak solution in $H^{2}(\mathbb{S})$.
Firstly, the Cauchy problem (2.6) can be rewritten as follows
$$
\begin{array}{l}
\left\{\begin{array}{l}
\partial_{t}u+u\partial_{x}u
 +\partial_{x}P=0,\quad t>0, ~x\in \mathbb{R},\\[5pt]
\Lambda_{\mu}^{4}P=2\mu(u)u-\frac{1}{2}(\partial_{x}^{2}u)^{2}
 -3\partial_{x}(\partial_{x}u\partial_{x}^{2}u),\\[5pt]
u(t,x+1)=u(t,x), \quad t\geq0, ~x\in \mathbb{R},\\[5pt]
u(0,x)=u_{0}(x), \quad x\in \mathbb{R}.
\end{array}
\right.
\end{array}
\eqno(5.1)
$$
Now we introduce the definition of a weak solution to the Cauchy problem (5.1).\\

\noindent\textbf{Definition 5.1.} We call
$u: \mathbb{R}_{+}\times \mathbb{S}\rightarrow \mathbb{R}$ an admissible global weak solution of the Cauchy problem (5.1) if\\
$(i)$ $u(t,x)\in C(\mathbb{R}_{+}; C^{1}(\mathbb{S}))\cap L^{\infty}(\mathbb{R}_{+}; H^{2}(\mathbb{S}))$
and
$$
\begin{array}{l}
\|\partial_{x}^{2}u(t,\cdot)\|_{L^{2}(\mathbb{S})}
\leq\|\partial_{x}^{2}u_{0}\|_{L^{2}(\mathbb{S})} \quad \mbox{for each}~ t>0.
\end{array}
\eqno(5.2)
$$
$(ii)$ $u(t,x)$ satisfies (5.1) in the sense of distributions and takes on the initial data pointwise.\\

The main result of this section is as follows.\\

\noindent\textbf{Theorem 5.1.} Let $p>2$. For any $u_{0}\in H^{2}(\mathbb{S})$ satisfying $\partial_{x}^{2}u_{0}\in L^{p}(\mathbb{S})$,
the Cauchy problem (5.1) has an admissible global weak solution in the sense of Definition 5.1.

\subsection{\textbf{Viscous Approximate Solutions}}

In this subsection, we construct the approximation solution sequence
$u_{\varepsilon}=u_{\varepsilon}(t,x)$. Hence, we consider the viscous problem of
(5.1) as follows
$$
\left\{\begin{array}{rl}
&\partial_{t}u_{\varepsilon}+u_{\varepsilon}\partial_{x}u_{\varepsilon}
+\partial_{x}P_{\varepsilon}=\varepsilon\partial_{x}^{2}u_{\varepsilon},
 \quad t>0,~x\in \mathbb{R},\\[3pt]
&\Lambda_{\mu}^{4}P_{\varepsilon}=2\mu(u_{\varepsilon})u_{\varepsilon}
 -\frac{1}{2}(\partial_{x}^{2}u_{\varepsilon})^{2}
 -3\partial_{x}(\partial_{x}u_{\varepsilon}\partial_{x}^{2}u_{\varepsilon}),
 \quad t\geq0,~x\in \mathbb{R},\\[3pt]
&u_{\varepsilon}(t,x+1)=u_{\varepsilon}(t,x), \quad t\geq0,~x\in \mathbb{R},\\[3pt]
&u_{\varepsilon}(0,x)=u_{\varepsilon,0}(x), \quad x\in \mathbb{R},
\end{array}
\right.
\eqno(5.3)
$$
where $u_{\varepsilon,0}(x)=(j_{\varepsilon}*u_{0})(x)$ and
$j_{\varepsilon}$ is defined in (3.1).
By Lemma 3.3, we have
$$
\begin{array}{l}
\|u_{\varepsilon,0}\|_{L^{2}(\mathbb{S})}\leq \|u_{0}\|_{L^{2}(\mathbb{S})},~
\|\partial_{x}^{2}u_{\varepsilon,0}\|_{L^{2}(\mathbb{S})}\leq \|\partial_{x}^{2}u_{0}\|_{L^{2}(\mathbb{S})},~
\|\partial_{x}^{2}u_{\varepsilon,0}\|_{L^{p}(\mathbb{S})}\leq \|\partial_{x}^{2}u_{0}\|_{L^{p}(\mathbb{S})}
\end{array}
\eqno(5.4)
$$
and $u_{\varepsilon,0}\rightarrow u_{0} ~ \mbox{in}~H^{2}(\mathbb{S})$.
Note that $\Lambda_{\mu}^{-4}\partial_{x}^{4}w
=w(x)-\int_{0}^{1}w(x)dx$. Differentiating the first equation of (5.3) three times with respect to $x$, one obtains
$$
\begin{array}{rl}
\partial_{t}\partial_{x}^{3}u_{\varepsilon}
+\partial_{x}u_{\varepsilon}\partial_{x}^{3}u_{\varepsilon}
+u_{\varepsilon}\partial_{x}^{4}u_{\varepsilon}+
2\mu(u_{\varepsilon})u_{\varepsilon}
 -\frac{1}{2}(\partial_{x}^{2}u_{\varepsilon})^{2}
-\varepsilon\partial_{x}^{5}u_{\varepsilon}=a_{\varepsilon}(t),
\end{array}
\eqno(5.5)
$$
where
$$
a_{\varepsilon}(t)
=-\frac{1}{2}\int_{\mathbb{S}}(\partial_{x}^{2}u_{\varepsilon})^{2}dx
  +2\mu(u_{\varepsilon})^{2}.
$$

\noindent\textbf{Lemma 5.1.}
Let $\varepsilon>0$ and $u_{\varepsilon,0}\in H^{s}(\mathbb{S})$, $s\geq 5$. Then
there exists a unique
$u_{\varepsilon}\in C(\mathbb{R}_{+}; H^{s}(\mathbb{S}))$
to (5.3). Moreover, for each $t\geq 0$ and $\varepsilon>0$, it holds that
$$
\begin{array}{rl}
\int_{\mathbb{S}}(\partial_{x}^{2} u_{\varepsilon})^{2}(t,x)dx
+2\varepsilon\int_{0}^{t}\int_{\mathbb{S}}
(\partial_{x}^{3} u_{\varepsilon})^{2}(s,x)dxds=\int_{\mathbb{S}}(\partial_{x}^{2} u_{\varepsilon,0})^{2}dx,
\end{array}
\eqno(5.6)
$$
and for each $\varepsilon>0$,
$$
\begin{array}{rl}
\|u_{\varepsilon}\|_{L^{\infty}(\mathbb{R}_{+}\times \mathbb{S})},~
\|\partial_{x}u_{\varepsilon}\|_{L^{\infty}(\mathbb{R}_{+}\times \mathbb{S})}
\leq \|u_{0}\|_{H^{2}(\mathbb{S})}.
\end{array}
$$

\noindent\textbf{Proof.}
First, following the standard argument for a nonlinear parabolic equation, one can obtain
the local well-posedness result that, for $u_{\varepsilon,0}\in H^{s}(\mathbb{S})$, $s\geq4$,
there exists a positive constant $T_{0}$ such that (5.3) has a unique solution
$u_{\varepsilon}=u_{\varepsilon}(t,x)\in C([0,T_{0}]; H^{s}(\mathbb{S}))\cap L^{2}([0,T_{0}]; H^{s+1}(\mathbb{S}))$. We denote the life span of the solution $u_{\varepsilon}(t,x)$ by $T$.
Multiplying (5.5) by $\partial_{x}u_{\varepsilon}$ and integrating over $\mathbb{S}$, we obtain
$$
-\frac{1}{2}\frac{d}{dt}\int_{\mathbb{S}}(\partial_{x}^{2} u_{\varepsilon})^{2}(t,x)dx
=\varepsilon\int_{\mathbb{S}}(\partial_{x}^{3} u_{\varepsilon})^{2}(t,x)dx.
$$
Then $(5.6)$ holds for all $0\leq t<T$.

Similar to the proof of Theorem 3.1, we can show that if
$\|\partial_{x}^{3}u_{\varepsilon}(t,\cdot)\|_{L^{\infty}(\mathbb{S})}
<\infty$, then the $H^{s}(\mathbb{S})$-norm of $u_{\varepsilon}(t,\cdot)$ does not blow up on $[0,T)$. Next, we prove $T=\infty$.
By Corollary 2.1, (5.4) and (5.6), we obtain that
$$
\begin{array}{rl}
\max_{x\in\mathbb{S}}(\partial_{x}u_{\varepsilon})^{2}(t,x)
&\leq \frac{1}{12}\int_{\mathbb{S}}(\partial_{x}^{2}u_{\varepsilon})^{2}(t,x)dx
\leq\frac{1}{12}\int_{\mathbb{S}}(\partial_{x}^{2} u_{\varepsilon,0})^{2}dx\\[5pt]
&\leq\frac{1}{12}\|\partial_{x}^{2}u_{0}\|^{2}_{L^{2}(\mathbb{S})}.
\end{array}
$$
This in turn implies that
$$
\begin{array}{l}
\|\partial_{x}u_{\varepsilon}\|_{L^{\infty}(\mathbb{S})}
\leq \frac{\sqrt{3}}{6}\|\partial_{x}^{2}u_{0}\|_{L^{2}(\mathbb{S})}.
\end{array}
$$
Note that
$
\int_{\mathbb{S}}(u_{\varepsilon}(t,x)-\mu_{\varepsilon,0})=0.
$
By Corollary 2.1,
$$
\begin{array}{rl}
\max_{x\in\mathbb{S}}(u_{\varepsilon}(t,x)-\mu_{\varepsilon,0})^{2}
\leq \frac{1}{12}\int_{\mathbb{S}}(\partial_{x}u_{\varepsilon})^{2}(t,x)dx
\leq \frac{1}{12}\|\partial_{x}u_{\varepsilon}(t,\cdot)\|_{L^{\infty}(\mathbb{S})}^{2}.
\end{array}
\eqno(5.7)
$$
Hence, we get
$$
\|u_{\varepsilon}\|_{L^{\infty}(\mathbb{S})}
\leq \frac{1}{12}\|\partial_{x}^{2}u_{0}\|_{L^{2}(\mathbb{S})}
+|\mu_{\varepsilon,0}|
\leq \frac{1}{12}\|\partial_{x}^{2}u_{0}\|_{L^{2}(\mathbb{S})}
+\|u_{0}\|_{L^{2}(\mathbb{S})}.
$$

Due to Corollary 2.1, we only need to derive an a priori estimate on
$\|\partial_{x}^{4}u_{\varepsilon}\|_{L^{2}(\mathbb{S})}$. In view of (5.5), we get
$$
\begin{array}{rl}
&\frac{1}{2}\frac{d}{dt}\int_{\mathbb{S}}(\partial_{x}^{4} u_{\varepsilon})^{2}(t,x)dx
+\varepsilon\int_{\mathbb{S}}(\partial_{x}^{5} u_{\varepsilon})^{2}(t,x)dx\\[5pt]
&=
-\frac{3}{2}\int_{\mathbb{S}}\partial_{x}u_{\varepsilon}(\partial_{x}^{4}u_{\varepsilon})^{2}dx
\leq
\frac{\sqrt{3}}{4}\|\partial_{x}^{2}u_{0}\|_{L^{2}(\mathbb{S})}
\|\partial_{x}^{4} u_{\varepsilon}\|^{2}_{L^{2}(\mathbb{S})},
\end{array}
$$
which implies that $\|\partial_{x}^{4} u_{\varepsilon}\|_{L^{2}(\mathbb{S})}$ is bounded for $t\in [0, T)$.
Since $\|\partial_{x}^{3}u_{\varepsilon}\|_{L^{\infty}(\mathbb{S})}
\leq \frac{\sqrt{3}}{6}\|\partial_{x}^{4} u_{\varepsilon}\|_{L^{2}(\mathbb{S})}$ by Corollary 2.1, we have $T=\infty$, which completes the proof of the lemma. \hfill $\Box$

\subsection{\textbf{Precompactness}}

In this subsection, we are ready to obtain the necessary compactness of the viscous approximation solutions $u_{\varepsilon}(t,x)$.

For convenience, we denote $P_{\varepsilon}=P_{1,\varepsilon}+P_{2,\varepsilon}$, where $P_{1,\varepsilon}, P_{2,\varepsilon}$ are defined by
$$
\begin{array}{rl}
&P_{1,\varepsilon}
=\Lambda_{\mu}^{-4}
[2\mu(u_{\varepsilon})u_{\varepsilon}-\frac{1}{2}(\partial_{x}^{2}u_{\varepsilon})^{2}],\\[5pt]
&P_{2,\varepsilon}
=
-3\partial_{x}\Lambda_{\mu}^{-4}(\partial_{x}u_{\varepsilon}\partial_{x}^{2}u_{\varepsilon}).
\end{array}
$$
Moreover,
$$
\begin{array}{rl}
\partial_{x}^{3}P_{\varepsilon}
=\partial_{x}^{3}P_{1,\varepsilon}+\partial_{x}^{3}P_{2,\varepsilon}
=\partial_{x}^{3}P_{1,\varepsilon}
-3\partial_{x}u_{\varepsilon}\partial_{x}^{2}u_{\varepsilon}.
\end{array}
$$

\noindent\textbf{Lemma 5.2.} Assume $u_{0}\in H^{2}(\mathbb{S})$. For each $t\geq 0$ and $\varepsilon>0$, the following inequalities hold
$$
\begin{array}{rl}
&\|P_{1,\varepsilon}(t, \cdot)\|_{W^{4,1}(\mathbb{S})},~
\|P_{1,\varepsilon}(t, \cdot)\|_{W^{3,\infty}(\mathbb{S})}
\leq C_{0}\|u_{0}\|_{H^{2}(\mathbb{S})}^{2},\\[5pt]
&\|P_{2,\varepsilon}(t, \cdot)\|_{W^{2,1}(\mathbb{S})},~
\|P_{2,\varepsilon}(t, \cdot)\|_{W^{2,\infty}(\mathbb{S})}
\leq C_{0}\|u_{0}\|_{H^{2}(\mathbb{S})}^{2},\\[5pt]
&\|P_{\varepsilon}(t, \cdot)\|_{W^{2,1}(\mathbb{S})},~
\|P_{\varepsilon}(t, \cdot)\|_{W^{2,\infty}(\mathbb{S})}
\leq C_{0}\|u_{0}\|_{H^{2}(\mathbb{S})}^{2},\\[5pt]
&\|\partial_{x}^{3}P_{\varepsilon}(t, \cdot)\|_{L^{1}(\mathbb{S})}
\leq C_{0}\|u_{0}\|_{H^{2}(\mathbb{S})}^{2}.
\end{array}
$$
Here and in what follows, we use $C_{0}$ to denote a generic positive constant, independent of $\varepsilon$, which may change from line to line.\\

\noindent\textbf{Proof.} For $\sigma=1$ or $\infty$, by $(2.2)$, we have
$$
\begin{array}{rl}
\|\partial_{x}P_{1,\varepsilon}(t, \cdot)\|_{L^{\sigma}(\mathbb{S})}
&\leq \left\|-\frac{x^{3}}{6}+\frac{x^{2}}{4}-\frac{x}{12}\right\|_{L^{\sigma}(\mathbb{S})}
\left|\int_{0}^{1}[2\mu(u_{\varepsilon})u_{\varepsilon}-\frac{1}{2}(\partial_{x}^{2}u_{\varepsilon})^{2}]dx\right|\\[5pt]
&\quad+\left\|-\frac{x^{2}}{2}+\frac{x}{2}-\frac{1}{12}\right\|_{L^{\sigma}(\mathbb{S})}
\left|\int_{0}^{1}\int_{0}^{x}[2\mu(u_{\varepsilon})u_{\varepsilon}-\frac{1}{2}(\partial_{y}^{2}u_{\varepsilon})^{2}]dydx\right|\\[5pt]
&\quad +\left\|-x+\frac{1}{2}\right\|_{L^{\sigma}(\mathbb{S})}
\left|\int_{0}^{1}\int_{0}^{x}\int_{0}^{y}[2\mu(u_{\varepsilon})u_{\varepsilon}-\frac{1}{2}(\partial_{z}^{2}u_{\varepsilon})^{2}]dzdydx\right|\\[5pt]
&\quad +\left|\int_{0}^{1}\int_{0}^{x}\int_{0}^{y}\int_{0}^{z}[2\mu(u_{\varepsilon})u_{\varepsilon}-\frac{1}{2}(\partial_{r}^{2}u_{\varepsilon})^{2}]drdzdydx\right|\\[5pt]
&\quad+\left\|\int_{0}^{x}\int_{0}^{z}\int_{0}^{r}[2\mu(u_{\varepsilon})u_{\varepsilon}-\frac{1}{2}(\partial_{s}^{2}u_{\varepsilon})^{2}]dsdrdz\right\|_{L^{\sigma}(\mathbb{S})}\\[5pt]
&\leq C_{0}\int_{0}^{1}|2\mu(u_{\varepsilon})u_{\varepsilon}-\frac{1}{2}(\partial_{x}^{2}u_{\varepsilon})^{2}|dx
\leq C_{0}\|u_{0}\|_{H^{2}(\mathbb{S})}^{2}
\end{array}
$$
and
$$
\begin{array}{rl}
\|\partial_{x}P_{2,\varepsilon}(t, \cdot)\|_{L^{\sigma}(\mathbb{S})}
&\leq 3\left\|-\frac{x^{2}}{2}+\frac{x}{2}-\frac{1}{12}\right\|_{L^{\sigma}(\mathbb{S})}
\left|\int_{0}^{1}\partial_{x}u_{\varepsilon}\partial_{x}^{2}u_{\varepsilon}dx\right|\\[5pt]
&\quad+3\left\|-x+\frac{1}{2}\right\|_{L^{\sigma}(\mathbb{S})}
\left|\int_{0}^{1}\int_{0}^{x}\partial_{y}u_{\varepsilon}\partial_{y}^{2}u_{\varepsilon}dydx\right|\\[5pt]
&\quad
+3\left|\int_{0}^{1}\int_{0}^{x}\int_{0}^{y}\partial_{z}u_{\varepsilon}\partial_{z}^{2}u_{\varepsilon}dzdydx\right|\\[5pt]
&\quad+3\left\|\int_{0}^{x}\int_{0}^{r}\partial_{s}u_{\varepsilon}\partial_{s}^{2}u_{\varepsilon}dsdr\right\|_{L^{\sigma}(\mathbb{S})}\\[5pt]
&\leq C_{0}\int_{0}^{1}|\partial_{x}u_{\varepsilon}\partial_{x}^{2}u_{\varepsilon}|dx
\leq C_{0}\|u_{0}\|_{H^{2}(\mathbb{S})}^{2}.
\end{array}
$$
The estimates of $\|\partial_{x}^{i}P_{1,\varepsilon}(t, \cdot)\|_{L^{\sigma}(\mathbb{S})}$ $(i=0,2,3)$, $\|\partial_{x}^{4}P_{1,\varepsilon}(t, \cdot)\|_{L^{1}(\mathbb{S})}$
and $\|\partial_{x}^{j}P_{2,\varepsilon}(t, \cdot)\|_{L^{\sigma}(\mathbb{S})}$ $(j=0,2)$ can be obtained similarly. Since $P_{\varepsilon}=P_{1,\varepsilon}+P_{2,\varepsilon}$, we can easily deduce the estimate of $\|P_{\varepsilon}(t, \cdot)\|_{W^{2,\sigma}(\mathbb{S})}$.

On the other hand, by Lemma 5.1, we have
$$
\begin{array}{rl}
\|\partial_{x}^{3}P_{\varepsilon}(t, \cdot)\|_{L^{1}(\mathbb{S})}
&=\|\partial_{x}^{3}P_{1,\varepsilon}(t, \cdot)
-3\partial_{x}u_{\varepsilon}(t, \cdot)\partial_{x}^{2}u_{\varepsilon}(t, \cdot)\|_{L^{1}(\mathbb{S})}\\[5pt]
&\leq \|\partial_{x}^{3}P_{1,\varepsilon}(t, \cdot)\|_{L^{1}(\mathbb{S})}
+3\|\partial_{x}u_{\varepsilon}(t, \cdot)\partial_{x}^{2}u_{\varepsilon}(t, \cdot)\|_{L^{1}(\mathbb{S})}\\[5pt]
&\leq \|\partial_{x}^{3}P_{1,\varepsilon}(t, \cdot)\|_{L^{1}(\mathbb{S})}
+3\|\partial_{x}u_{\varepsilon}(t, \cdot)\|_{L^{\infty}(\mathbb{S})}\|\partial_{x}^{2}u_{\varepsilon}(t, \cdot)\|_{L^{2}(\mathbb{S})}\\[5pt]
&\leq C_{0}\|u_{0}\|_{H^{2}(\mathbb{S})}^{2},
\end{array}
$$
which completes the proof. \hfill $\Box$\\

Next we turn to estimates of time derivatives.\\

\noindent\textbf{Lemma 5.3.} Assume $u_{0}\in H^{2}(\mathbb{S})$. For each $T, t>0$ and $0<\varepsilon<1$,
the following inequalities hold
$$
\begin{array}{rl}
&\|\partial_{t}u_{\varepsilon}(t, \cdot)\|_{L^{2}(\mathbb{S})}
\leq C_{0}\|u_{0}\|_{H^{2}(\mathbb{S})}^{2}
+\|u_{0}\|_{H^{2}(\mathbb{S})}, \\[5pt]
&\|\partial_{t}\partial_{x}u_{\varepsilon}\|_{L^{2}([0, T]\times\mathbb{S})}
\leq C_{0}\sqrt{T}\|u_{0}\|_{H^{2}(\mathbb{S})}^{2}
+\frac{\sqrt{2}}{2} \|u_{0}\|_{H^{2}(\mathbb{S})}, \\[5pt]
&\|\partial_{t}\partial_{x}^{3}P_{1,\varepsilon}\|_{L^{1}([0, T]\times\mathbb{S})}
\leq C_{0}[T\|u_{0}\|_{H^{2}(\mathbb{S})}^{3}+(T+1)\|u_{0}\|_{H^{2}(\mathbb{S})}^{2}].
\end{array}
$$

\noindent\textbf{Proof.} By the first equation of (5.3) and Lemmas 5.1-5.2, we have
$$
\begin{array}{rl}
&\|\partial_{t}u_{\varepsilon}(t, \cdot)\|_{L^{2}(\mathbb{S})}\\[3pt]
&\leq \|u_{\varepsilon}(t, \cdot)\partial_{x}u_{\varepsilon}(t, \cdot)\|_{L^{2}(\mathbb{S})}
+\|\partial_{x}P_{\varepsilon}(t, \cdot)\|_{L^{2}(\mathbb{S})}
+\varepsilon\|\partial_{x}^{2}u_{\varepsilon}(t, \cdot)\|_{L^{2}(\mathbb{S})}\\[3pt]
&\leq \|u_{\varepsilon}(t, \cdot)\|_{L^{\infty}(\mathbb{S})}\|\partial_{x}u_{\varepsilon}(t, \cdot)\|_{L^{\infty}(\mathbb{S})}
+\|\partial_{x}P_{\varepsilon}(t, \cdot)\|_{L^{\infty}(\mathbb{S})}
+\varepsilon\|\partial_{x}^{2}u_{\varepsilon}(t, \cdot)\|_{L^{2}(\mathbb{S})}\\[3pt]
&\leq C_{0}\|u_{0}\|_{H^{2}(\mathbb{S})}^{2}
+\varepsilon \|u_{0}\|_{H^{2}(\mathbb{S})}
\leq C_{0}\|u_{0}\|_{H^{2}(\mathbb{S})}^{2}
+\|u_{0}\|_{H^{2}(\mathbb{S})}.
\end{array}
$$
Differentiating the first equation of (5.3) with respect to $x$, one obtains
$$
\begin{array}{rl}
\partial_{t}\partial_{x}u_{\varepsilon}
+u_{\varepsilon}\partial_{x}^{2}u_{\varepsilon}
+(\partial_{x}u_{\varepsilon})^{2}
+\partial_{x}^{2}P_{\varepsilon}
=\varepsilon\partial_{x}^{3}u_{\varepsilon}.
\end{array}
$$
Thus,
$$
\begin{array}{rl}
\|\partial_{t}\partial_{x}u_{\varepsilon}\|_{L^{2}([0, T]\times\mathbb{S})}
&\leq \|u_{\varepsilon}\partial_{x}^{2}u_{\varepsilon}\|_{L^{2}([0, T]\times\mathbb{S})}
+\|\partial_{x}u_{\varepsilon}\|_{L^{4}([0, T]\times\mathbb{S})}^{2}\\[3pt]
&\quad +\|\partial_{x}^{2}P_{\varepsilon}\|_{L^{2}([0, T]\times\mathbb{S})}
+\varepsilon\|\partial_{x}^{3}u_{\varepsilon}\|_{L^{2}([0, T]\times\mathbb{S})}\\[3pt]
&\leq \|u_{\varepsilon}\|_{L^{\infty}([0, T]\times\mathbb{S})}\|\partial_{x}^{2}u_{\varepsilon}\|_{L^{2}([0, T]\times\mathbb{S})}
+\sqrt{T}\|\partial_{x}u_{\varepsilon}\|_{L^{\infty}([0, T]\times\mathbb{S})}^{2}\\[3pt]
&\quad +\sqrt{T}\|\partial_{x}^{2}P_{\varepsilon}\|_{L^{\infty}([0, T]\times\mathbb{S})}
+\varepsilon\|\partial_{x}^{3}u_{\varepsilon}\|_{L^{2}([0, T]\times\mathbb{S})}\\[3pt]
&\leq C_{0}\sqrt{T}\|u_{0}\|_{H^{2}(\mathbb{S})}^{2}
+\frac{\sqrt{2}}{2} \|u_{0}\|_{H^{2}(\mathbb{S})}.
\end{array}
$$
Moreover, by the definition of $P_{1,\varepsilon}$ and (2.1), we know
$$
\begin{array}{rl}
\partial_{t}\partial_{x}^{3}P_{1,\varepsilon}
&=\partial_{t}\partial_{x}^{3}\Lambda_{\mu}^{-4}
[2\mu(u_{\varepsilon})u_{\varepsilon}-\frac{1}{2}(\partial_{x}^{2}u_{\varepsilon})^{2}]\\[3pt]
&=(-x+\frac{1}{2})\int_{0}^{1}(2\mu(u_{\varepsilon})\partial_{t}u_{\varepsilon}
-\partial_{x}^{2}u_{\varepsilon}\partial_{t}\partial_{x}^{2}u_{\varepsilon})dx\\[3pt]
&\quad -\int_{0}^{1}\int_{0}^{x}(2\mu(u_{\varepsilon})\partial_{t}u_{\varepsilon}
-\partial_{y}^{2}u_{\varepsilon}\partial_{t}\partial_{y}^{2}u_{\varepsilon})dydx\\[3pt]
&\quad +\int_{0}^{x}(2\mu(u_{\varepsilon})\partial_{t}u_{\varepsilon}
-\partial_{y}^{2}u_{\varepsilon}\partial_{t}\partial_{y}^{2}u_{\varepsilon})dy.
\end{array}
\eqno(5.8)
$$
Differentiating the first equation of (5.3) with respect to $x$ two times, we have
$$
\begin{array}{rl}
\partial_{t}\partial_{x}^{2}u_{\varepsilon}
+u_{\varepsilon}\partial_{x}^{3}u_{\varepsilon}
+\partial_{x}^{3}P_{1,\varepsilon}
=\varepsilon\partial_{x}^{4}u_{\varepsilon},
\end{array}
$$
and then
$$
\begin{array}{rl}
&\partial_{x}^{2}u_{\varepsilon}\partial_{t}\partial_{x}^{2}u_{\varepsilon}\\[3pt]
&=-\partial_{x}^{2}u_{\varepsilon}(u_{\varepsilon}\partial_{x}^{3}u_{\varepsilon}
+\partial_{x}^{3}P_{1,\varepsilon}
-\varepsilon\partial_{x}^{4}u_{\varepsilon})\\[3pt]
&=-\frac{1}{2}\partial_{x}(u_{\varepsilon}(\partial_{x}^{2}u_{\varepsilon})^{2})
+\frac{1}{2}\partial_{x}u_{\varepsilon}(\partial_{x}^{2}u_{\varepsilon})^{2}
-\partial_{x}^{2}u_{\varepsilon}\partial_{x}^{3}P_{1,\varepsilon}
+\varepsilon\partial_{x}(\partial_{x}^{2}u_{\varepsilon}\partial_{x}^{3}u_{\varepsilon})
-\varepsilon(\partial_{x}^{3}u_{\varepsilon})^{2}.
\end{array}
$$
Thus,
$$
\begin{array}{rl}
&(x-\frac{1}{2})\int_{0}^{1}\partial_{x}^{2}u_{\varepsilon}\partial_{t}\partial_{x}^{2}u_{\varepsilon}dx
+\int_{0}^{1}\int_{0}^{x}\partial_{y}^{2}u_{\varepsilon}\partial_{t}\partial_{y}^{2}u_{\varepsilon}dydx
-\int_{0}^{x}\partial_{y}^{2}u_{\varepsilon}\partial_{t}\partial_{y}^{2}u_{\varepsilon}dy\\[5pt]
&=(x-\frac{1}{2})\int_{0}^{1}[\frac{1}{2}\partial_{x}u_{\varepsilon}(\partial_{x}^{2}u_{\varepsilon})^{2}
-\partial_{x}^{2}u_{\varepsilon}\partial_{x}^{3}P_{1,\varepsilon}
-\varepsilon(\partial_{x}^{3}u_{\varepsilon})^{2}]dx\\[5pt]
&\quad +\int_{0}^{1}\int_{0}^{x}[\frac{1}{2}\partial_{y}u_{\varepsilon}(\partial_{y}^{2}u_{\varepsilon})^{2}
-\partial_{y}^{2}u_{\varepsilon}\partial_{y}^{3}P_{1,\varepsilon}
-\varepsilon(\partial_{y}^{3}u_{\varepsilon})^{2}]dydx\\[5pt]
&\quad -\int_{0}^{x}[\frac{1}{2}\partial_{y}u_{\varepsilon}(\partial_{y}^{2}u_{\varepsilon})^{2}
-\partial_{y}^{2}u_{\varepsilon}\partial_{y}^{3}P_{1,\varepsilon}
-\varepsilon(\partial_{y}^{3}u_{\varepsilon})^{2}]dy\\[5pt]
&\quad
-\frac{1}{2}\int_{0}^{1}u_{\varepsilon}(\partial_{x}^{2}u_{\varepsilon})^{2}dx
+\frac{1}{2}u_{\varepsilon}(\partial_{x}^{2}u_{\varepsilon})^{2}
-\varepsilon \partial_{x}^{2}u_{\varepsilon}\partial_{x}^{3}u_{\varepsilon}.
\end{array}
\eqno(5.9)
$$
From Lemmas 5.1-5.2, (5.8) and (5.9), we get
$$
\begin{array}{rl}
&\|\partial_{t}\partial_{x}^{3}P_{1,\varepsilon}\|_{L^{1}([0, T]\times\mathbb{S})}\\[3pt]
&\leq C_{0}(\int_{0}^{T}\int_{0}^{1}|\mu(u_{\varepsilon})\partial_{t}u_{\varepsilon}|dxdt
+\int_{0}^{T}\int_{0}^{1}|\frac{1}{2}\partial_{x}u_{\varepsilon}(\partial_{x}^{2}u_{\varepsilon})^{2}
-\partial_{x}^{2}u_{\varepsilon}\partial_{x}^{3}P_{1,\varepsilon}
-\varepsilon(\partial_{x}^{3}u_{\varepsilon})^{2}|dxdt\\[3pt]
&\quad +\int_{0}^{T}\int_{0}^{1}|u_{\varepsilon}(\partial_{x}^{2}u_{\varepsilon})^{2}|dxdt
+\int_{0}^{T}\int_{0}^{1}|\varepsilon \partial_{x}^{2}u_{\varepsilon}\partial_{x}^{3}u_{\varepsilon}|dxdt)\\[3pt]
&\leq C_{0}(\sqrt{T}|\mu(u_{\varepsilon})|\|\partial_{t}u_{\varepsilon}\|_{L^{2}([0, T]\times\mathbb{S})}
+\|\partial_{x}u_{\varepsilon}\|_{L^{\infty}([0, T]\times\mathbb{S})}
\|\partial_{x}^{2}u_{\varepsilon}\|_{L^{2}([0, T]\times\mathbb{S})}^{2}\\[3pt]
&\quad +\sqrt{T}\|\partial_{x}^{2}u_{\varepsilon}\|_{L^{2}([0, T]\times\mathbb{S})}
\|\partial_{x}^{3}P_{1,\varepsilon}\|_{L^{\infty}([0, T]\times\mathbb{S})}
+\varepsilon\|\partial_{x}^{3}u_{\varepsilon}\|_{L^{2}([0, T]\times\mathbb{S})}^{2}\\[3pt]
&\quad +\|u_{\varepsilon}\|_{L^{\infty}([0, T]\times\mathbb{S})}
\|\partial_{x}^{2}u_{\varepsilon}\|_{L^{2}([0, T]\times\mathbb{S})}^{2}
+\varepsilon\|\partial_{x}^{2}u_{\varepsilon}\|_{L^{2}([0, T]\times\mathbb{S})}
\|\partial_{x}^{3}u_{\varepsilon}\|_{L^{2}([0, T]\times\mathbb{S})})\\[3pt]
&\leq C_{0}[T\|u_{0}\|_{H^{2}(\mathbb{S})}^{3}+(T+1)\|u_{0}\|_{H^{2}(\mathbb{S})}^{2}],
\end{array}
$$
which completes the proof of Lemma 5.3. \hfill $\Box$\\

\noindent\textbf{Lemma 5.4.} Let $u_{0}\in H^{2}(\mathbb{S})$ and $\partial_{x}^{2}u_{0}\in L^{p}(\mathbb{S})$ for some $p>2$. Then the following inequality holds
$$
\begin{array}{rl}
\|\partial_{x}^{2}u_{\varepsilon}(t, \cdot)\|_{L^{p}(\mathbb{S})}
\leq e^{\|u_{0}\|_{H^{2}(\mathbb{S})}t}\|\partial_{x}^{2}u_{0}\|_{L^{p}(\mathbb{S})}
+pC_{0}\|u_{0}\|_{H^{2}(\mathbb{S})}(e^{\|u_{0}\|_{H^{2}(\mathbb{S})}t}-1).
\end{array}
$$

\noindent\textbf{Proof.}
Denote $q_{\varepsilon}:=\partial_{x}^{2}u_{\varepsilon}$, then $q_{\varepsilon}$ satisfies
$$
\begin{array}{rl}
\partial_{t}q_{\varepsilon}
+u_{\varepsilon}\partial_{x}q_{\varepsilon}
+\partial_{x}^{3}P_{1,\varepsilon}
=\varepsilon\partial_{x}^{2}q_{\varepsilon}.
\end{array}
\eqno(5.10)
$$
Multiplying the above equation by $pq_{\varepsilon}|q_{\varepsilon}|^{p-2}$, we have
$$
\begin{array}{rl}
\partial_{t}(|q_{\varepsilon}|^{p})
+u_{\varepsilon}\partial_{x}(|q_{\varepsilon}|^{p})
+pq_{\varepsilon}|q_{\varepsilon}|^{p-2}\partial_{x}^{3}P_{1,\varepsilon}
&=\varepsilon pq_{\varepsilon}|q_{\varepsilon}|^{p-2}\partial_{x}^{2}q_{\varepsilon}\\[3pt]
&=\varepsilon\partial_{x}^{2}(|q_{\varepsilon}|^{p})
-\varepsilon p(p-1)|q_{\varepsilon}|^{p-2}(\partial_{x}q_{\varepsilon})^{2}.
\end{array}
$$
By Lemmas 5.1-5.2, we know
$$
\begin{array}{rl}
\frac{d}{dt}\int_{\mathbb{S}}|q_{\varepsilon}|^{p}dx
&\leq \int_{\mathbb{S}}\partial_{x}u_{\varepsilon}|q_{\varepsilon}|^{p}dx
+p\int_{\mathbb{S}}|q_{\varepsilon}|^{p-1}|\partial_{x}^{3}P_{1,\varepsilon}|dx\\[3pt]
&\leq \|\partial_{x}u_{\varepsilon}\|_{L^{\infty}(\mathbb{S})}
\int_{\mathbb{S}}|q_{\varepsilon}|^{p}dx
+p\|\partial_{x}^{3}P_{1,\varepsilon}\|_{L^{p}(\mathbb{S})}
\|q_{\varepsilon}\|_{L^{p}(\mathbb{S})}^{p-1}\\[3pt]
&\leq \|u_{0}\|_{H^{2}(\mathbb{S})}\int_{\mathbb{S}}|q_{\varepsilon}|^{p}dx
+pC_{0}\|u_{0}\|_{H^{2}(\mathbb{S})}^{2}\|q_{\varepsilon}\|_{L^{p}(\mathbb{S})}^{p-1}.
\end{array}
$$
Note that
$$
\begin{array}{rl}
\frac{d}{dt}\int_{\mathbb{S}}|q_{\varepsilon}|^{p}dx
=p\|q_{\varepsilon}\|_{L^{p}(\mathbb{S})}^{p-1}\frac{d}{dt}\|q_{\varepsilon}\|_{L^{p}(\mathbb{S})}.
\end{array}
$$
Thus,
$$
\begin{array}{rl}
\frac{d}{dt}\|q_{\varepsilon}\|_{L^{p}(\mathbb{S})}
\leq \|u_{0}\|_{H^{2}(\mathbb{S})}\|q_{\varepsilon}\|_{L^{p}(\mathbb{S})}
+pC_{0}\|u_{0}\|_{H^{2}(\mathbb{S})}^{2}.
\end{array}
$$
The Gronwall inequality implies the desired result.
\hfill $\Box$\\

\noindent\textbf{Lemma 5.5.} Let $u_{0}\in H^{2}(\mathbb{S})$ and $\partial_{x}^{2}u_{0}\in L^{p}(\mathbb{S})$ for some $p>2$. There exist a positive sequence $\{\varepsilon_{k}\}_{k\in \mathbb{N}}$ decreasing to zero and three functions $u\in L^{\infty}(\mathbb{R}_{+}; H^{2}(\mathbb{S}))\cap H^{1}(\mathbb{R}_{+}\times\mathbb{S})\subseteq C(\mathbb{R}_{+}; C^{1}(\mathbb{S}))$ for each $T>0$, $P\in L^{\infty}(\mathbb{R}_{+}; W^{2, \infty}(\mathbb{S}))$
and $\widetilde{P}\in L^{\infty}(\mathbb{R}_{+}; W^{1, 1}(\mathbb{S})\cap L^{\infty}(\mathbb{S}))$ such that
$$
\begin{array}{rl}
&u_{\varepsilon_{k}}\rightharpoonup u \quad
\mbox{weakly~in}~H^{1}([0, T]\times \mathbb{S})~\mbox{for~each}~T\geq 0;\\[3pt]
&u_{\varepsilon_{k}}\rightarrow u \quad
\mbox{strongly~in}~L_{loc}^{\infty}(\mathbb{R}_{+}; H^{1}(\mathbb{S}));\\[3pt]
&P_{\varepsilon_{k}}\rightharpoonup P \quad
\mbox{weakly~in}~L_{loc}^{\sigma}(\mathbb{R}_{+}\times\mathbb{S})~\mbox{for~each}~1< \sigma<\infty;\\[3pt]
&\partial_{x}^{3}P_{1,\varepsilon_{k}}\rightarrow \widetilde{P} \quad
\mbox{strongly~in}~L_{loc}^{\sigma}(\mathbb{R}_{+}\times\mathbb{S})~\mbox{for~each}~1\leq \sigma<\infty.
\end{array}
$$

\noindent\textbf{Proof.}
Due to Lemmas 5.1 and 5.3, we have that
$$
\begin{array}{c}
\{u_{\varepsilon}\}_{\varepsilon}~
\mbox{is~uniformly~bounded~in}~L^{\infty}(\mathbb{R}_{+}; H^{2}(\mathbb{S})),\\[3pt]
\{\partial_{t}u_{\varepsilon}\}_{\varepsilon}~
\mbox{is~uniformly~bounded~in}~L^{2}([0, T]; H^{1}(\mathbb{S}))~\mbox{for~each}~T> 0.
\end{array}
$$
In particular, $\{u_{\varepsilon}\}_{\varepsilon}$ is uniformly bounded in
$H^{1}([0, T]\times\mathbb{S})$ and then we have $u_{\varepsilon_{k}}\rightharpoonup u$
weakly in $H^{1}([0, T]\times \mathbb{S})$.
Moreover, Using the fact $H^{2}(\mathbb{S})\Subset H^{1}(\mathbb{S})\Subset L^{2}(\mathbb{S})$ and Corollary 4 in \cite{simon87}, we know that $u_{\varepsilon_{k}}\rightarrow u$ strongly in $L_{loc}^{\infty}(\mathbb{R}_{+}; H^{1}(\mathbb{S}))$.

Due to Lemma 5.2, we obtain that
$$
\begin{array}{c}
\{P_{\varepsilon}\}_{\varepsilon}~
\mbox{is~uniformly~bounded~in}~L^{\infty}(\mathbb{R}_{+}; W^{2, \infty}(\mathbb{S})).
\end{array}
$$
In particular, $\{P_{\varepsilon}\}_{\varepsilon}$ is uniformly bounded in
$L^{\sigma}([0, T]\times\mathbb{S})$ with $1< \sigma<\infty$ and then we have
$P_{\varepsilon_{k}}\rightharpoonup P$ weakly in $L_{loc}^{\sigma}(\mathbb{R}_{+}\times\mathbb{S})$ for each $1< \sigma<\infty$.

Due to Lemmas 5.2 and 5.3,
$$
\begin{array}{c}
\{\partial_{x}^{3}P_{1,\varepsilon}\}_{\varepsilon}~
\mbox{is~uniformly~bounded~in}~L^{\infty}(\mathbb{R}_{+}; W^{1, 1}(\mathbb{S})\cap L^{\infty}(\mathbb{S})),\\[3pt]
\{\partial_{t}\partial_{x}^{3}P_{1,\varepsilon}\}_{\varepsilon}~
\mbox{is~uniformly~bounded~in}~L^{1}([0, T]\times\mathbb{S})~\mbox{for~each}~T> 0.
\end{array}
$$
Using the fact $W^{1, 1}(\mathbb{S})\Subset L^{\sigma}(\mathbb{S})\subset L^{1}(\mathbb{S})$, $1\leq\sigma<\infty$ and Corollary 4 in \cite{simon87},
we know that $\partial_{x}^{3}P_{1,\varepsilon_{k}}\rightarrow \widetilde{P}$
strongly in $L_{loc}^{\sigma}(\mathbb{R}_{+}\times\mathbb{S})$ for each $1\leq\sigma<\infty$. \hfill $\Box$

\subsection{\textbf{Existence of solutions}}

From Lemmas 5.1 and 5.4, we can deduce that there exist two functions
$q\in L_{loc}^{\rho}(\mathbb{R}_{+}\times\mathbb{S})$
and $\overline{q^{2}}\in L_{loc}^{r}(\mathbb{R}_{+}\times\mathbb{S})$ such that
$$
\begin{array}{l}
q_{\varepsilon_{k}}\rightharpoonup q \quad
\mbox{in}~L_{loc}^{\rho}(\mathbb{R}_{+}\times\mathbb{S}),
\quad
q^{2}_{\varepsilon_{k}}\rightharpoonup \overline{q^{2}} \quad
\mbox{in}~L_{loc}^{r}(\mathbb{R}_{+}\times\mathbb{S}),
\end{array}
\eqno(5.11)
$$
for every $1<\rho<p$ and $1<r<\frac{p}{2}$. Moreover,
$$
q^{2}(t,x)\leq \overline{q^{2}}(t,x),\quad a.e.~(t,x)\in \mathbb{R}_{+}\times\mathbb{S}.
$$

In view of (5.11), we conclude that for any $\eta\in C^{1}(\mathbb{R})$ with $\eta^{\prime}$
bounded, Lipschitz continuous on $\mathbb{R}$, $\eta(0)=0$ and any $1<\rho<p$, we have
$$
\eta(q_{\varepsilon_{k}})\rightharpoonup \overline{\eta(q)}
\quad \mbox{in}~L_{loc}^{\rho}(\mathbb{R}_{+}\times\mathbb{S}).
$$
Here and in what follows, we use overbars to denote weak limits in spaces to be understood
from the context.\\

\noindent\textbf{Lemma 5.6.}
The following inequality holds in the sense of distributions
$$
\begin{array}{rl}
&\int_{\mathbb{S}}\left(\overline{(q_{+})^{2}}-(q_{+})^{2}\right)dx
\leq \int_{0}^{t}\int_{\mathbb{S}}\partial_{x}u\left(\overline{(q_{+})^{2}}-(q_{+})^{2}\right)dtdx
-2\int_{0}^{t}\int_{\mathbb{S}}\widetilde{P}(\overline{q_{+}}-q_{+})dtdx.
\end{array}
\eqno(5.12)
$$

\noindent\textbf{Proof.}
The proof is divided into the following three steps.\\
\emph{Step 1.}
Taking $\xi\in C^{2}(\mathbb{R})$ convex
and multiplying (5.10) by $\xi^{\prime}(q_{\varepsilon_{k}})$, we have
$$
\begin{array}{rl}
&\partial_{t}\xi(q_{\varepsilon_{k}})
+\partial_{x}(u_{\varepsilon_{k}}\xi(q_{\varepsilon_{k}}))
-\xi(q_{\varepsilon_{k}})\partial_{x}u_{\varepsilon_{k}}
+\xi^{\prime}(q_{\varepsilon_{k}})\partial_{x}^{3}P_{1,\varepsilon_{k}}\\[3pt]
&=\varepsilon_{k}\partial_{x}^{2}\xi(q_{\varepsilon_{k}})
-\varepsilon_{k}\xi^{\prime\prime}(q_{\varepsilon_{k}})(\partial_{x}q_{\varepsilon_{k}})^{2}
\leq\varepsilon_{k}\partial_{x}^{2}\xi(q_{\varepsilon_{k}}).
\end{array}
\eqno(5.13)
$$
In particular, we can use the entropy $q\mapsto (q_{+})^{2}/2$ and get
$$
\begin{array}{rl}
&\partial_{t}\frac{(q_{\varepsilon_{k},+})^{2}}{2}
+\partial_{x}
\left(u_{\varepsilon_{k}}\frac{(q_{\varepsilon_{k},+})^{2}}{2}\right)
-\frac{(q_{\varepsilon_{k},+})^{2}}{2}\partial_{x}u_{\varepsilon_{k}}
+q_{\varepsilon_{k},+}\partial_{x}^{3}P_{1,\varepsilon_{k}}
\leq \varepsilon_{k}\partial_{x}^{2}\frac{(q_{\varepsilon_{k},+})^{2}}{2}.
\end{array}
$$
Letting $k\rightarrow \infty$, we have
$$
\begin{array}{l}
\partial_{t}\frac{\overline{(q_{+})^{2}}}{2}
+\partial_{x}
\left(u\frac{\overline{(q_{+})^{2}}}{2}\right)
-\frac{\overline{(q_{+})^{2}}}{2}\partial_{x}u
+\widetilde{P}\overline{q_{+}}\leq 0.
\end{array}
\eqno(5.14)
$$
\emph{Step 2.} Using (5.10), Lemma 5.5 and (5.11), letting $\varepsilon\rightarrow 0$, we have
$$
\begin{array}{l}
\partial_{t}q+\partial_{x}\left(uq\right)
-q\partial_{x}u+\widetilde{P}
=0. \end{array}
\eqno(5.15)
$$
Denote $q^{\delta}(t,x):=(q(t,\cdot)\ast j_{\delta})(x)$, where $j_{\delta}$
is the Friedrichs mollifier defined in (3.1).
According to Lemma II.1 in \cite{dili89}, it follows from (5.15) that $q^{\delta}$ solves
$$
\begin{array}{rl}
\partial_{t}q^{\delta}+u\partial_{x}q^{\delta}+\widetilde{P}*j_{\delta}
=\tau_{\delta},
\end{array}
\eqno(5.16)
$$
where the error $\tau_{\delta}$ tends to zero in $L_{loc}^{1}(\mathbb{R}_{+}\times\mathbb{S})$.
Multiplying (5.16) by $\eta^{\prime}(q^{\delta})$, we have
$$
\begin{array}{rl}
\partial_{t}\eta(q^{\delta})
+\partial_{x}\left(u\eta(q^{\delta})\right)
-\eta(q^{\delta})\partial_{x}u
+(\widetilde{P}*j_{\delta})\eta^{\prime}(q^{\delta})
=\tau_{\delta}\eta^{\prime}(q^{\delta}).
\end{array}
\eqno(5.17)
$$
Using the boundedness of $\eta$ and $\eta^{\prime}$ and sending $\delta\rightarrow 0$
in (5.17), we obtain
$$
\begin{array}{rl}
&\partial_{t}\eta(q)+\partial_{x}\left(u\eta(q)\right)
-\eta(q)\partial_{x}u+\widetilde{P}\eta^{\prime}(q)
=0.
\end{array}
\eqno(5.18)
$$
In particular, we can use the entropy $\eta_{R}^{+}(\xi):=\eta_{R}(\xi)\chi_{[0,\infty)}(\xi)$,
where $R>0$, $\chi_{E}$ is the characteristic function in set $E$ and
$$
\begin{array}{rl}
\eta_{R}(\xi)
:=\left\{\begin{array}{l}
\frac{1}{2}\xi^{2}, \quad \mbox{if}~|\xi|\leq R, \\[5pt]
R|\xi|-\frac{1}{2}R^{2}, \quad \mbox{if}~|\xi|>R,
\end{array}
\right.
\end{array}
$$
and then get
$$
\begin{array}{rl}
\partial_{t}\eta_{R}^{+}(q)+\partial_{x}\left(u\eta_{R}^{+}(q)\right)
-\eta_{R}^{+}(q)\partial_{x}u
+\widetilde{P}(\eta_{R}^{+})^{\prime}(q)
=0.
\end{array}
\eqno(5.19)
$$
\emph{Step 3.} Subtracting (5.19) from (5.14), we get
$$
\begin{array}{rl}
&\partial_{t}\left(\frac{\overline{(q_{+})^{2}}}{2}-\eta_{R}^{+}(q)\right)
+\partial_{x}\left(u(\frac{\overline{(q_{+})^{2}}}{2}-\eta_{R}^{+}(q))\right)\\[3pt]
&\quad-\partial_{x}u\left(\frac{\overline{(q_{+})^{2}}}{2}-\eta_{R}^{+}(q)\right)
+\widetilde{P}\left(\overline{q_{+}}-(\eta_{R}^{+})^{\prime}(q)\right)
\leq 0.
\end{array}
\eqno(5.20)
$$

Integrating (5.20) over $(0,t)\times \mathbb{S}$, we get
$$
\begin{array}{rl}
&\int_{\mathbb{S}}\left(\frac{\overline{(q_{+})^{2}}}{2}-\eta_{R}^{+}(q)\right)(t,x)dx
-\int_{\mathbb{S}}\left(\frac{\overline{(q_{+})^{2}}}{2}-\eta_{R}^{+}(q)\right)(0,x)dx\\[5pt]
&\leq\int_{0}^{t}\int_{\mathbb{S}}\partial_{x}u\left(\frac{\overline{(q_{+})^{2}}}{2}-\eta_{R}^{+}(q)\right)dtdx
-\int_{0}^{t}\int_{\mathbb{S}}\widetilde{P}\left(\overline{q_{+}}-(\eta_{R}^{+})^{\prime}(q)\right)dtdx.
\end{array}
\eqno(5.21)
$$
Since
$$
\begin{array}{rl}
\eta_{R}^{+}(q)=\frac{1}{2}q_{+}^{2}-\frac{1}{2}(R-q)^{2}\chi_{(R,\infty)}(q),
\quad (\eta_{R}^{+})^{\prime}(q)=q_{+}+(R-q)\chi_{(R,\infty)}(q),
\end{array}
$$
we have, as $R\rightarrow\infty$,
$$
\begin{array}{rl}
\eta_{R}^{+}(q)(t,x)\rightarrow \frac{1}{2}q_{+}^{2}(t,x),\quad
(\eta_{R}^{+})^{\prime}(q)(t,x)\rightarrow q_{+}(t,x) \quad\mbox{a.e.~in}~[0,t)\times \mathbb{S}
\end{array}
$$
and for any $R\geq1$
$$
\begin{array}{rl}
(R-q)^{2}\chi_{(R,\infty)}(q)\leq (1-q)^{2}\chi_{(1,\infty)}(q),\quad
(R-q)\chi_{(R,\infty)}(q)\leq (1-q)\chi_{(1,\infty)}(q).
\end{array}
$$
By applying the Dominate Convergence Theorem in (5.21), we know
$$
\begin{array}{rl}
&\int_{\mathbb{S}}\left(\frac{\overline{(q_{+})^{2}}}{2}-\frac{q_{+}^{2}}{2}\right)(t,x)dx
-\int_{\mathbb{S}}\left(\frac{\overline{(q_{+})^{2}}}{2}-\frac{q_{+}^{2}}{2}\right)(0,x)dx\\[5pt]
&\leq\int_{0}^{t}\int_{\mathbb{S}}\partial_{x}u\left(\frac{\overline{(q_{+})^{2}}}{2}-\frac{q_{+}^{2}}{2}\right)dtdx
-\int_{0}^{t}\int_{\mathbb{S}}\widetilde{P}\left(\overline{q_{+}}-q_{+}\right)dtdx.
\end{array}
$$
Similar to the arguments in \cite{chkar06,xz00}, there holds
$$
\begin{array}{l}
\lim_{t\rightarrow 0^{+}}\int_{\mathbb{S}}q^{2}(t,x)dx
=\lim_{t\rightarrow 0^{+}}\int_{\mathbb{S}}\overline{q^{2}}(t,x)dx
=\int_{\mathbb{S}}(\partial_{x}^{2}u_{0})^{2}(x)dx.
\end{array}
\eqno(5.22)
$$
Since the functions $\xi_{+}^{2}$ and $\xi_{-}^{2}$ are convex in $\xi$, we have
$$
\begin{array}{l}
0\leq\overline{(q_{+})^{2}}-q_{+}^{2}
\leq\overline{(q)^{2}}-q^{2},
\end{array}
$$
and then
$$
\begin{array}{l}
\int_{\mathbb{S}}\left(\frac{\overline{(q_{+})^{2}}}{2}-\frac{q_{+}^{2}}{2}\right)(0,x)dx
=0,
\end{array}
$$
which completes the proof.
\hfill $\Box$\\

By using the entropy $\eta_{R}^{-}(\xi):=\eta_{R}(\xi)\chi_{[-\infty,0]}(\xi)$,
where $\eta_{R}(\xi)$ is defined in the proof of Lemma 5.6, we can obtain the following lemma.\\

\noindent\textbf{Lemma 5.7.} For any $t>0$ and any $R>0$,
$$
\begin{array}{rl}
&\int_{\mathbb{S}}\left(\overline{\eta_{R}^{-}(q)}-\eta_{R}^{-}(q)\right)dx
\leq
\int_{0}^{t}\int_{\mathbb{S}}\partial_{x}u\left(\overline{\eta_{R}^{-}(q)}-\eta_{R}^{-}(q)\right)dtdx\\[5pt]
&\quad-\int_{0}^{t}\int_{\mathbb{S}}\widetilde{P}
\left(\overline{(\eta_{R}^{-})^{\prime}(q)}-(\eta_{R}^{-})^{\prime}(q)\right)dtdx. \end{array}
\eqno(5.23)
$$

\noindent\textbf{Proof.}
Letting $k\rightarrow \infty$ in (5.18), we have
$$
\begin{array}{rl}
\partial_{t}\overline{\eta(q)}+\partial_{x}\left(u\overline{\eta(q)}\right)
-\overline{\eta(q)}\partial_{x}u+\widetilde{P}\overline{\eta^{\prime}(q)}
\leq 0.
\end{array}
\eqno(5.24)
$$
Subtracting (5.18) from (5.24) and using the entropy $\eta_{R}^{-}(\xi)$,
we have
$$
\begin{array}{rl}
&\partial_{t}\left(\overline{\eta_{R}^{-}(q)}-\eta_{R}^{-}(q)\right)
+\partial_{x}
\left(u\left(\overline{\eta_{R}^{-}(q)}-\eta_{R}^{-}(q)\right)\right)\\[3pt]
&\leq \partial_{x}u\left(\overline{\eta_{R}^{-}(q)}-\eta_{R}^{-}(q)\right)
-\widetilde{P}\left(\overline{(\eta_{R}^{-})^{\prime}(q)}-(\eta_{R}^{-})^{\prime}(q)\right).
\end{array}
\eqno(5.25)
$$
Similar to the arguments in \cite{chkar06,xz00}, for each $R>0$ we have
$$
\begin{array}{l}
\lim_{t\rightarrow 0^{+}}
\int_{\mathbb{S}}\left(\overline{\eta_{R}^{-}(q)}(t,x)-\eta_{R}^{-}(q)(t,x)\right)dx=0. \end{array}
$$
Integrating (5.25) over $(0,t)\times \mathbb{S}$, we get (5.23). \hfill $\Box$\\

\noindent\textbf{Lemma 5.8.} There holds $\overline{q^{2}}=q^{2}$, a.e. on $\mathbb{R}_{+}\times \mathbb{S}$.\\

\noindent\textbf{Proof.}
Adding (5.12) and (5.23) yields
$$
\begin{array}{rl}
&\int_{\mathbb{S}}\frac{1}{2}\left(\overline{(q_{+})^{2}}-(q_{+})^{2}\right)+
\left(\overline{\eta_{R}^{-}(q)}-\eta_{R}^{-}(q)\right)dx\\[5pt]
&\leq\int_{0}^{t}\int_{\mathbb{S}}\partial_{x}u\left(\frac{1}{2}\overline{(q_{+})^{2}}-\frac{1}{2}(q_{+})^{2}
+\overline{\eta_{R}^{-}(q)}-\eta_{R}^{-}(q)\right)dtdx\\[5pt]
&\quad-\int_{0}^{t}\int_{\mathbb{S}}\widetilde{P}
\left(\overline{q_{+}}-q_{+}
+\overline{(\eta_{R}^{-})^{\prime}(q)}-(\eta_{R}^{-})^{\prime}(q)\right)dtdx. \end{array}
\eqno(5.26)
$$
By the definition of $\eta_{R}^{-}(\xi)$,
$$
\begin{array}{rl}
&(\overline{q_{+}}-q_{+})
+\left(\overline{(\eta_{R}^{-})^{\prime}(q)}-(\eta_{R}^{-})^{\prime}(q)\right)\\[5pt]
&=(R+q)\chi_{(-\infty,-R)}(q)-\overline{(R+q)\chi_{(-\infty,-R)}(q)}.
\end{array}
$$
Similar as Step 3 in the proof of Lemma 5.6, we can deduce the following inequality by applying the Dominate Convergence Theorem in (5.26)
$$
\begin{array}{rl}
0
&\leq\int_{\mathbb{S}}\frac{1}{2}\left(\overline{(q_{+})^{2}}-(q_{+})^{2}\right)+
\frac{1}{2}\left(\overline{(q_{-})^{2}}-(q_{-})^{2}\right)dx\\[3pt]
&\leq\int_{0}^{t}\int_{\mathbb{S}}\partial_{x}u\left(\frac{1}{2}\overline{(q_{+})^{2}}-\frac{1}{2}(q_{+})^{2}+
\frac{1}{2}\overline{(q_{-})^{2}}-\frac{1}{2}(q_{-})^{2}\right)dtdx.
\end{array}
$$
Since $\partial_{x}u$ is uniformly bounded, using (5.22) and Gronwall's inequality, we conclude that
$$
0\leq\int_{\mathbb{S}}\left(\overline{q^{2}}-q^{2}\right)dx\leq 0\quad \mbox{for~any}~t>0,
$$
which completes the proof.  \hfill $\Box$\\

\noindent\textbf{Proof of Theorem 5.1.}
Let $u(t,x)$ be the limit of the viscous approximation solutions $u_{\varepsilon}(t,x)$
as $\varepsilon\rightarrow 0$. It then follows from Lemmas 5.1 and 5.5 that
$u(t,x)\in C(\mathbb{R}_{+}; C^{1}(\mathbb{S}))\cap L^{\infty}(\mathbb{R}_{+}; H^{2}(\mathbb{S}))$
and (5.2) holds.

Let $\mu_{t,x}(\lambda)$ be the Young measure
associated with $\{q_{\varepsilon}\}=\{\partial_{x}^{2}u_{\varepsilon}\}$, see more details in Lemma 4.2 of \cite{xz00}.
According to Lemma 5.8, we have $\mu_{t,x}(\lambda)=\delta_{\overline{q(t,x)}}(\lambda)$ a.e.
$(t,x)\in \mathbb{R}_{+}\times \mathbb{S}$, then
$$
\begin{array}{l}
q_{\varepsilon}=\partial_{x}^{2}u_{\varepsilon}\rightarrow q=\partial_{x}^{2}u
\quad\mbox{in}~L^{2}_{loc}(\mathbb{R}_{+}\times \mathbb{S}).
\end{array}
\eqno(5.27)
$$
Taking $\varepsilon\rightarrow 0$ in (5.3), one finds from (5.27)
and Lemma 5.5 that $u(t,x)$ is an admissible weak solution to (5.1).
This completes the proof of Theorem 5.1. \hfill $\Box$

\section{Peaked Solutions}

Recall first that the single peakons for $\mu$-Camassa-Holm equation are given by $u(t,x)=\frac{12c}{13}g_{\mu}(x-ct)$ in \cite{lmt10}, where
$g_{\mu}(x)=\frac{1}{2}(x-[x]-\frac{1}{2})^{2}+\frac{23}{24}$ is the Green's function of the operator $(\mu-\partial_{x}^{2})^{-1}$ with $[x]$ being the largest integer part of $x$.
It is worth mentioning that the modified $\mu$-Camassa-Holm equations introduced in \cite{qfliu14,qfl14} also admit single peakons represented by the Green's function $g_{\mu}(x)$. In this section, we show the existence of single peakon solutions to (1.1).\\

\noindent\textbf{Theorem 6.1.} For any $c>0$, (1.1) admits the peaked periodic-one traveling-wave
solutions $u_{c}(t,x)=\phi(\xi)$, $\xi=x-ct$, where
$$
\begin{array}{l}
\phi(\xi)
=\frac{720}{721}c\left[-\frac{1}{24}\left((\xi-[\xi]-\frac{1}{2})^{2}-\frac{1}{4}\right)^{2}+\frac{721}{720}\right].
\end{array}
$$

\noindent\textbf{Proof.} Motivated by the forms of periodic peakons for the $\mu$-Camassa-Holm equation \cite{lmt10} and modified $\mu$-Camassa-Holm equations \cite{qfliu14,qfl14}, we assume that the periodic peakon of (1.1) is given by
$$
\begin{array}{l}
u_{c}(t, x)
=a\left[-\frac{1}{24}\left((\xi-[\xi]-\frac{1}{2})^{2}-\frac{1}{4}\right)^{2}+\frac{721}{720}\right].
\end{array}
$$
According to the definition of weak solutions, $u_{c}(t, x)$ satisfies the following equation
$$
\begin{array}{rl}
\Sigma_{j=1}^{4}I_{j}
&:=\int_{0}^{T}\int_{\mathbb{S}}u_{c,t}\varphi dxdt
+\int_{0}^{T}\int_{\mathbb{S}}u_{c}u_{c,x}\varphi dxdt\\[5pt]
&\quad+\int_{0}^{T}\int_{\mathbb{S}}g_{x}*[2\mu(u_{c})u_{c}-\frac{1}{2}u_{c,xx}^{2}]\varphi dxdt
-\int_{0}^{T}\int_{\mathbb{S}}3g_{xx}*(u_{c,x}u_{c,xx})\varphi dxdt\\[5pt]
&=0,
\end{array}
$$
for some $T>0$ and any test function $\varphi(t,x)\in C_{c}^{\infty}([0, T)\times \mathbb{S})$,
where $g(x)$ is the Green function defined in Section 2. One can obtain that
$$
\begin{array}{rl}
\mu(u_{c})
&=a\int_{0}^{ct}\left[-\frac{1}{24}\left((x-ct+\frac{1}{2})^{2}-\frac{1}{4}\right)^{2}+\frac{721}{720}\right]dx\\[5pt]
&\quad+a\int_{ct}^{1}\left[-\frac{1}{24}\left((x-ct-\frac{1}{2})^{2}-\frac{1}{4}\right)^{2}+\frac{721}{720}\right]dx\\[5pt]
&=a.
\end{array}
$$

To evaluate $I_{j}$, $j=1,2,3,4$, we need to consider two cases: $(i)$ $x>ct$
and $x\leq ct$.

For $x>ct$, we get
$$
\begin{array}{rl}
&g_{x}*[2\mu(u_{c})u_{c}-\frac{1}{2}u_{c,xx}^{2}]\\[5pt]
&=a^{2}\int_{\mathbb{S}}\left(-\frac{1}{6}(x-y-[x-y]-\frac{1}{2})^{3}
+\frac{1}{24}(x-y-[x-y]-\frac{1}{2})\right)\\[5pt]
&\quad \times\left(-\frac{5}{24}(y-ct-[y-ct]-\frac{1}{2})^{4}+\frac{1}{16}(y-ct-[y-ct]-\frac{1}{2})^{2}
+\frac{11501}{5760}\right)dy\\[5pt]
&=a^{2}\int_{0}^{ct}\left(-\frac{1}{6}(x-y-\frac{1}{2})^{3}
+\frac{1}{24}(x-y-\frac{1}{2})\right) \left(-\frac{5}{24}(y-ct+\frac{1}{2})^{4}+\frac{1}{16}(y-ct+\frac{1}{2})^{2}
+\frac{11501}{5760}\right)dy\\[5pt]
&\quad+a^{2}\int_{ct}^{x}\left(-\frac{1}{6}(x-y-\frac{1}{2})^{3}
+\frac{1}{24}(x-y-\frac{1}{2})\right) \left(-\frac{5}{24}(y-ct-\frac{1}{2})^{4}+\frac{1}{16}(y-ct-\frac{1}{2})^{2}
+\frac{11501}{5760}\right)dy\\[5pt]
&\quad+a^{2}\int_{x}^{1}\left(-\frac{1}{6}(x-y+\frac{1}{2})^{3}
+\frac{1}{24}(x-y+\frac{1}{2})\right) \left(-\frac{5}{24}(y-ct-\frac{1}{2})^{4}+\frac{1}{16}(y-ct-\frac{1}{2})^{2}
+\frac{11501}{5760}\right)dy
\end{array}
$$
and
$$
\begin{array}{rl}
&g_{xx}*(u_{c,x}u_{c,xx})\\[5pt]
&=a^{2}\int_{\mathbb{S}}\left(-\frac{1}{2}(x-y-[x-y]-\frac{1}{2})^{2}
+\frac{1}{24}\right)\\[5pt]
&\quad \times\left(\frac{1}{12}(y-ct-[y-ct]-\frac{1}{2})^{5}-\frac{1}{36}(y-ct-[y-ct]-\frac{1}{2})^{3}
+\frac{1}{576}(y-ct-[y-ct]-\frac{1}{2})\right)dy\\[5pt]
&=a^{2}\int_{0}^{ct}\left(-\frac{1}{2}(x-y-\frac{1}{2})^{2}
+\frac{1}{24}\right)
\left(\frac{1}{12}(y-ct+\frac{1}{2})^{5}-\frac{1}{36}(y-ct+\frac{1}{2})^{3}
+\frac{1}{576}(y-ct+\frac{1}{2})\right)dy\\[5pt]
&\quad+a^{2}\int_{ct}^{x}\left(-\frac{1}{2}(x-y-\frac{1}{2})^{2}
+\frac{1}{24}\right)
\left(\frac{1}{12}(y-ct-\frac{1}{2})^{5}-\frac{1}{36}(y-ct-\frac{1}{2})^{3}
+\frac{1}{576}(y-ct-\frac{1}{2})\right)dy\\[5pt]
&\quad+a^{2}\int_{x}^{1}\left(-\frac{1}{2}(x-y+\frac{1}{2})^{2}
+\frac{1}{24}\right)
\left(\frac{1}{12}(y-ct-\frac{1}{2})^{5}-\frac{1}{36}(y-ct-\frac{1}{2})^{3}
+\frac{1}{576}(y-ct-\frac{1}{2})\right)dy.
\end{array}
$$
By calculations, we have
$$
\begin{array}{rl}
&g_{x}*[2\mu(u_{c})u_{c}-\frac{1}{2}u_{c,xx}^{2}]
-3g_{xx}*(u_{c,x}u_{c,xx})\\[5pt]
&=a^{2}[-\frac{1}{144}(x-ct)^{7}+\frac{7}{288}(x-ct)^{6}
-\frac{1}{32}(x-ct)^{5}
+\frac{5}{288}(x-ct)^{4}
-\frac{1}{288}(x-ct)^{3}].
\end{array}
$$
Since
$$
\begin{array}{rl}
u_{c,t}=ac\left[\frac{1}{6}(x-ct)^{3}-\frac{1}{4}(x-ct)^{2}+\frac{1}{12}(x-ct)\right]
\end{array}
$$
and
$$
\begin{array}{rl}
u_{c}u_{c,x}
&=a^{2}[\frac{1}{144}(x-ct)^{7}-\frac{7}{288}(x-ct)^{6}
+\frac{1}{32}(x-ct)^{5}-\frac{5}{288}(x-ct)^{4}\\[3pt]
&\qquad-\frac{353}{2160}(x-ct)^{3}+\frac{721}{2880}(x-ct)^{2}
-\frac{721}{8640}(x-ct)],
\end{array}
$$
we have
$$
\begin{array}{rl}
\Sigma_{j=1}^{4}I_{j}
=\int_{0}^{T}\int_{\mathbb{S}} a(c-\frac{721}{720}a)\left[\frac{1}{6}(x-ct)^{3}
-\frac{1}{4}(x-ct)^{2}+\frac{1}{12}(x-ct)\right]\varphi dxdt.
\end{array}
$$
Similarly, for $x\leq ct$, we have
$$
\begin{array}{c}
u_{c,t}=ac\left[\frac{1}{6}(x-ct)^{3}+\frac{1}{4}(x-ct)^{2}+\frac{1}{12}(x-ct)\right]\\[3pt]
u_{c}u_{c,x}=a^{2}[\frac{1}{144}(x-ct)^{7}+\frac{7}{288}(x-ct)^{6}
+\frac{1}{32}(x-ct)^{5}+\frac{5}{288}(x-ct)^{4}\\[3pt]
\qquad-\frac{353}{2160}(x-ct)^{3}-\frac{721}{2880}(x-ct)^{2}
-\frac{721}{8640}(x-ct)],
\end{array}
$$
and
$$
\begin{array}{rl}
&g_{x}*[2\mu(u_{c})u_{c}-\frac{1}{2}u_{c,xx}^{2}]
-3g_{xx}*(u_{c,x}u_{c,xx})\\[5pt]
&=a^{2}[-\frac{1}{144}(x-ct)^{7}
-\frac{7}{288}(x-ct)^{6}
-\frac{1}{32}(x-ct)^{5}
-\frac{5}{288}(x-ct)^{4}
-\frac{1}{288}(x-ct)^{3}].
\end{array}
$$
Thus,
$$
\begin{array}{rl}
\Sigma_{j=1}^{4}I_{j}
=\int_{0}^{T}\int_{\mathbb{S}} a(c-\frac{721}{720}a)\left[\frac{1}{6}(x-ct)^{3}
+\frac{1}{4}(x-ct)^{2}+\frac{1}{12}(x-ct)\right]\varphi dxdt.
\end{array}
$$
Since $\varphi$ is arbitrary, both cases imply $a$ satisfies $c-\frac{721}{720}a=0$, which completes the proof. \hfill $\Box$

\section*{Acknowledgments}
Wang's work was supported by the Fundamental Research Funds for the Central Universities.
Li's work was supported by the NSFC (No:11571057).
Qiao's work was partially supported by the President's Endowed Professorship program of the University of Texas system.

\label{}





\bibliographystyle{model3-num-names}
\bibliography{<your-bib-database>}



\end{document}